\documentclass{aa} 
\usepackage{graphicx}
\usepackage{txfonts}
\usepackage{textcomp}
\usepackage[dvipsnames]{xcolor}
\usepackage{mathrsfs}
\usepackage[colorlinks=true, citecolor=blue]{hyperref}

\newcommand{\rossbi}{\textit{RoSSBi3D}}
\newcommand{\vr}{\vec{r}}

\newcommand{\et}{\Vec{e}_\theta}
\newcommand{\ez}{\Vec{e}_z}

\begin{document} 


\title{RoSSBi3D: a 3D and bi-fluid code for protoplanetary discs}

\author{S. Rendon Restrepo \inst{1,2,3}
\thanks{\email{srendon@aip.de}}
\and
P. Barge \inst{1}
\thanks{\email{pierre.barge@lam.fr}}
\and
R. Vavrik \inst{4}
\thanks{\email{radim.vavrik@vsb.cz}}
}
          
\institute{Aix Marseille Univ, CNRS, CNES, LAM, Marseille, France
\and 
Aix Marseille Univ, CNRS, Centrale Marseille, IRPHE, Marseille, France
\and
Leibniz-Institut für Astrophysik Potsdam (AIP), Potsdam, Germany
\and
IT4Innovations, VSB – Technical University of Ostrava, 17. listopadu 2172/15, 708 00 Ostrava-Poruba, Czech Republic
}


 
\abstract
{The diversity of structures recently observed in protoplanetary discs (PPDs) with the new generation of high-resolution instruments have made the challenging questions that planet-formation models have to answer more acute. 
The challenge is on the theoretical side and in the numerical one, with the need to significantly improve the codes' performances and stretch the limit of PPD simulations. 
Multi-physics, fast, accurate, high-resolution, modular and reliable 3D codes are needed to understand the shape of PPD and the observed features as well as to explore the mechanisms at work.
}
{We present \rossbi{} as the 3D extension of the 2D code Rotating-System Simulations for Bi-fluids (\emph{RoSSBi}), a pure hydrodynamic code, which was specifically developed to study the evolution of PPDs.
Being described in detail, this is a new code, even if based on the 2D version.
We explain its architecture and specificity but also its performances against test cases.
}
{This grid-based, FORTRAN code solves the fully compressible inviscid continuity, Euler and energy conservation equations for an ideal gas in non-homentropic conditions and for pressureless particles in a fluid approximation.
It is a finite volume code which is second order in time and accounts for discontinuities thanks to an exact Riemann solver. 
The spatial scheme accounts for the equilibrium solution and is improved thanks to parabolic interpolation.
The code is developed in 3D and structured for high-performance parallelism.
}
{The optimised version of the code works on high-performance computers with excellent scalability. 
We checked its reliability against a 2D analogue of the sod shock tube test and a series of tests.
We release this code under the terms of the CeCILL2 Licence and make it publicly available.
}
{}


\maketitle
%

\section{Introduction}

Protoplanetary discs (PPDs) are mainly composed of molecular gas in which are embedded solid particles of different sizes.
An accurate description of both phases' dynamics could help to understand the aspect of PPDs at different wavelengths \citep{HH30_hubble,villenave_2020,villenave_2022} and the observed substructures \citep{2018_DSHARP,2018ApJ...860..124D,2018ApJ...869L..50P} with, as an ultimate goal, the will to unravel a global picture of planet formation.
Analytical tools remain fruitful in this quest since they led to the discovery of complex phenomenons possibly occurring in PPDs, such as the streaming instability \citep{2005_youdin,2007_youdin,2007_johansen}, vertical shear instability (VSI) \citep{nelson_gressel_2013,2014A&A...572A..77S,manger2020,manger2021}, gaseous vortices generation \citep{1999_lovelace}, dust trapping by vortices \citep{Barge2005}, baroclinic instabilities \citep{2003_klahr,2004_klahr} etc.
However, these tools reach their limits when accounting for a fine time evolution and global description of PPDs where multi-physic aspects are at interplay.
Therefore numerical experiments have become increasingly important in PPDs studies, and more generally in Astrophysics.
Progress in this field was made possible thanks to the development of accurate numerical methods, increasingly optimised algorithms, and the advent of multi-core machines, which have allowed access to unprecedented detail and multi-physics through the development of high-performance computing (HPC).

\rossbi{} is the result of the wish to develop a dedicated numerical tool suited to study the evolution of PPDs and adapted to their specifics. 
Among them, shearing by the gravitational potential of the star, vertical stratification and the intricate coupling of dust and gas.
In this code, we assumed that the gas, compressible and inviscid, is coupled through aerodynamic forces to solid particles which are described statistically as a fluid (or a family of different fluids) in which the intrinsic pressure is neglected.
This approximation remains relevant as far as dilution is strong, otherwise the inelastic collision between particles should be taken into account contributing to a pressure term.
Thus, in the limit of a diluted medium hypothesis, the problem is to solve a system of Euler's equations for the gas and the particles which are coupled to one another and to the energy equation.
This 3D extension of \emph{RoSSBi} \citep{survi2013} came from the necessity to provide a full description of the evolution of PPDs which requires to account for the disc thickness. 
It is indeed important to describe at best the physical evolution of the PPDs avoiding the constraining 2D approximation that could either enhance problems raised by the thin disc approximation, such as vortices destruction by dust back-reaction \citep{2021ApJ...913...92R}, or neglect physical aspects raised by vertical stratification, such as the VSI.

After assessing the three main discretisation methods used in fluid mechanics, we choose to keep the finite volume method (FVM), used in prior prototype versions, to solve the set of hyperbolic equations governing the evolution of an inviscid PPD.
Indeed, the finite difference method (FDM), well adapted for solving the partial differential equations related to Euler's equations, could provide incorrect results in the presence of rarefaction and shock waves, which could be encountered in PPDs \footnote{But this could be circumvent introducing an artificial bulk viscosity as proposed by \citet{von_neumann_richtmyer_1950}.}.
In order to account for discontinuities without introducing an artificial viscosity, the equations have to be solved under their most primitive form, which makes both the finite element method (FEM) and the FVM the natural choices for solving numerically the equations governing gas dynamics.
Indeed, both methods are well adapted for solving equations written in a conservative form and, as a consequence, are preferred choices for fluid dynamics problems.
The difficulty of the variational formalism and the problem reformulation under the weak form, intrinsic to FEM, lead us to naturally prefer the FVM.
The FVM has the advantage to be easier to understand and implement, and it requires only to evaluate fluxes at cells boundaries, which makes it a robust method for non-linear conservation laws \citep{2009..book..wendt,Toro_inbook} .

The goal of this paper is to present the new 3D \rossbi{} code, highlight its improvements in terms of performance, and grant this code with a proper reference for future scientific publications.
We attach great importance to the reliability of this code, thanks to a validation procedure with relevant test cases absent in previous versions and its scalability, which permits an efficient use of computational resources.
Even if it could not be reflected in this paper, the code was cleared, simplified, restructured, and highly documented.
\rossbi{} is part of a long story of 2D prototype codes, which led to a number of publications \citep{Inaba2005,Inaba2006,Richard2013,survi2013,Richard2016,Barge2016,2017A&A...605A.122B,rendonbarge_2022}.
The detailed legacy and numerical scheme evolution of these prototypes, on which several people have worked, is summarised in the code wiki.
This last version is based on the 2D code \emph{RoSSBi}, implemented by \citet{survi2013}, which was then upgraded in 2018, by Centre de donnéeS Astrophysiques de Marseille (CeSAM), with an elementary parallelisation using the MPI standard.
This code is written in the \href{https://fortran-lang.org/}{FORTRAN 90} programming language and requires the use of \href{https://www.hdfgroup.org/solutions/hdf5/}{HDF5} and \href{https://www.fftw.org/}{FFTW3} libraries.
\rossbi{} is open source and released under the terms of the \href{http://www.cecill.info/index.fr.html}{CeCILL2 Licence}.
The code can be downloaded from its public repository, \href{git@gitlab.lam.fr:srendon/rossbi3d.git}{git@gitlab.lam.fr:srendon/rossbi3d.git}.
The visualisation tools are not available in the repository but could be happily shared upon request.

The paper is organised in the following way: 
in Sect. \ref{sec:numerical resolution}, we explain the structure of the code, the numerical scheme, the choice of the boundary conditions, and the current supported options.
In Sect. \ref{sec:improving performances on parallel architectures}, we discuss in detail the performances of \rossbi{}.
Sect. \ref{sec:test cases} is devoted to benchmark  \rossbi{} numerical scheme with test cases. 
In Sect. \ref{sec:next steps and possible improvements}, we discuss additional steps in the development of our code and possible ways to improve it.
Finally, in Sect. \ref{sec:conclusion} we conclude.
Appendices mainly concerns aspects of the 3D stationary solution, required for balancing the numerical scheme, the conservative form of different terms, and the visualisation tools.


\section{Numerical resolution and code structure}\label{sec:numerical resolution}


This section is in the continuation of \citet[in french]{survi2013} work and devoted to introduce the numerical methods used in \rossbi{}.
We begin by recalling the description of the integral equations governing the gas dynamics.

\subsection{Conservative form of the equations}

Our code was developed assuming the gas is completely inviscid to conserve the equilibrium solution, to treat compressibility at best, and to respect the supersonic character of the Keplerian flow.
These conditions correspond to the ones expected in an ideal dead zone of PPDs.
Thus, in this outline, the equations governing gas dynamics in a 3D PPD are Euler's equation coupled with the conservation of mass and energy; they read: 
\begin{equation}\label{Eq:governing equations conservative}
\begin{array}{llll}
\partial_t \displaystyle\iiint\limits_{\mathcal{V}} \rho \, d\tau                                             \!\!\! & + 
\displaystyle\oiint\limits_{S} \rho \vec{V} \cdot \vec{dA}                                       \!\!\! & = & \!\!\!
\displaystyle\iiint\limits_{\mathcal{V}} \mathcal{S}_\rho \, d\tau                                                                                                         \\[4ex]
\partial_t \displaystyle\iiint\limits_{\mathcal{V}} \rho \vec{V} \, d\tau                                     \!\!\! & +   
\displaystyle\oiint\limits_{S} \left(\rho \vec{V} \otimes \vec{V} + \mathbb{I} P \right) \vec{dA}\!\!\! & = & \!\!\!
\displaystyle\iiint\limits_{\mathcal{V}} \vec{f}_V d\tau                                                     \\[4ex]
\partial_t \displaystyle\iiint\limits_{\mathcal{V}} \mathcal{E} \, d\tau                                                  \!\!\! & +  
\displaystyle\oiint\limits_{S} \left( \mathcal{E} + P \right) \vec{V} \cdot \vec{dA}                         \!\!\! & = & \!\!\!
\displaystyle\iiint\limits_{\mathcal{V}} \vec{V} \cdot \vec{f}_V d\tau \\[4ex]
   & & + & \!\!\! \displaystyle\iiint\limits_{\mathcal{V}}  \mathcal{S}_\mathcal{E} \, d\tau
\end{array}
\end{equation}
where $\mathcal{E} = \frac{\rho \vec{V}^2}{2} + \frac{P}{\gamma-1}$ is the gas energy and $\vec{f}_V$ are all volume forces applied to the gas.
$\mathcal{S}_\rho$ refers to source or sink terms in the continuity equation while $\mathcal{S}_\mathcal{E}$ stands for heat transfer by conduction, radiation, or cooling.
The variable $S$ is the surface enclosing the volume $\mathcal{V}$.
In current version of \rossbi{}, the volume forces applied to the gas are the central's object gravitation,  $\vec{f}_{star}=-\frac{GM_{\odot}\,\rho}{\left( r^2+z^2 \right)^{\frac{3}{2}}} \left(r \vec{e}_r + z \vec{e}_z\right)$, aerodynamic forces with solid particles, self-gravity (SG)\footnote{SG is not yet supported in the 3D version} and the indirect potential due to a shift of the barycenter of mass, since the frame is centered on the star's position.
The equations governing dust dynamics are slightly different, and it was preferred to devote the entire Sect. \ref{subsec:dust treatment} to solid particles.
The conservative form provided by Eqs. \ref{Eq:governing equations conservative} is attractive if the solution is discontinuous, which occurs when the flow is inviscid.
This is one of the primary reasons that led us to use the FVM that we describe in the next section.

\subsection{Numerical scheme: finite volume method}

We summarise the set of Eqs.  \ref{Eq:governing equations conservative} under the compact form: 
\begin{equation}\label{Eq:conservative form}
\partial_t \displaystyle\iiint\limits_{\mathcal{V}} E_p \, d\tau + 
\displaystyle\oiint\limits_{S} \vec{F}_p \cdot \vec{dA} = 
\displaystyle\iiint\limits_{\mathcal{V}} S_p \, d\tau
\end{equation}
where $E_p$, $F_p$, and $S_p$ are the $p^{th}$ component of the Euler variables, flux, and source vectors.
The expression of these quantities in cylindrical coordinates is provided in Appendix \ref{app:conservative form variables}.
We use a cell-centred formulation that is variables are defined by their mean value over finite volumes.
This implies that triple integrals are simply approximated on elementary volumes as follows: $\iiint\limits_{\mathcal{V}_{i,j,k}} f(E_p)\, d \tau \approx \overline{f}(E_p) \, \mathcal{V}_{i,j,k}$ where $\overline{f}(E_p)$ is the mean value of function $f$ in the elementary cell of respective volume $\mathcal{V}_{i,j,k}$. 
For conciseness, we will use the notation $f(E_p)=\overline{f}(E_p)$ throughout the rest of this paper.
On the other hand, the incoming fluxes are decomposed as elementary fluxes in each cell surface element.
Considering the above statements, the flux and source terms of Eq. \ref{Eq:conservative form} can be rewritten as: 
\begin{equation}
\begin{array}{ccccc}
\mathscr{S}_p(t, E_q)& =     &\displaystyle\iiint\limits_{\mathcal{V}_{i,j,k}} \mathcal{S}_p \, d\tau 
                     &\approx& S_p(E_q) \, \mathcal{V}_{i,j,k} \\
\mathscr{F}_p(t, E_q)&=&\displaystyle\oiint\limits_{S} \vec{F}_p \cdot \vec{dA} 
                     &\approx& \displaystyle\sum\limits_{S} A_s \vec{F}_p(E_q) \cdot \vec{n}_s  
\end{array}
\end{equation}
where $\vec{n}_s$ is the outward unit normal vector to surface $A_s$. 
Accounting for the above approximations, the conservative form of the evolution Eqs. \ref{Eq:conservative form} can be simply written:
\begin{equation}\label{Eq:compact form}
\mathcal{V}_{i,j,k} \, \partial_t  E_p = \mathscr{G}_p(t, E_q) \quad \mbox{with} \quad \mathscr{G}_p(t, E_q) = \mathscr{S}_p(E_q)-\mathscr{F}_p(t, E_q)
\end{equation}
The 2D and 3D versions of \rossbi{} solve the above equations in a polar and cylindrical coordinate system, respectively.

For the sake of conciseness, in next sections we will only treat the 1D case, which is easily generalised to three dimensions.

\subsection{Mesh}\label{subsec:mesh}

In the current version, we use a uniform structured grid in the azimuth and vertical directions while logarithmic in the radial direction.
The choice of a logarithmic mesh in the radial direction stems from the 2D code. 
Indeed, a logarithmic mesh is necessary for computing SG thanks to Fast Fourier methods \citep{2008_baruteau, survi2013, rendonbarge_2023}.
Even if the logarithmic cutting is unnecessary in the 3D version, we kept it in order to save the effort during the coding stage.

\subsection{Boundary conditions and ghost cells}

In our code, we connected the numerical domain to the stationary flow in isothermal equilibrium (see Eq. \ref{Eq: equilibrium solution}) thanks to a set of 2 ghost cells in each direction.
A sudden change in the flow at the domain boundaries could lead to waves reflection, which could affect the computation domain, and even instabilities can arise from an improper choice of BC.
Therefore, in the radial direction, we apply wake-killing and sheared BC in order to avoid (i) spurious reflections which can perturb the computation domain and (ii) the triggering of Papaloizou-Pringle like instabilities in the disc inner edge \citep{1984MNRAS.208..721P, 1985MNRAS.213P...7G}. 
We naturally use periodic BC in the azimuthal direction to close the physical domain.
Finally, motivated by the work of \citet{2012MNRAS.422.2399M,2013MNRAS.428..190L}, we implemented two different kinds of BC for the vertical direction: outflow BC or sheared damped BC.


\subsection{Time evolution and scheme stability}

In order to match high accuracy and reasonable computation effort, we use an explicit, second order Runge-Kutta Method (RKM2) time-stepping scheme for time integration.
Crudely, this RKM2 is based on the first prediction of a differential equation solution which permits performing a correction and thus increases the accuracy of the final solution.
The most general form of the explicit RKM2 applied to our problem is written: 
\begin{equation}
\begin{array}{ccl}
E^{n+1}_p & = & E^{n}_p                                                    \\ [4pt]
          & + & a \, \mathcal{V} \, \Delta t \, \mathscr{G}_p(t_n,E^{n}_q) \\ [4pt]
          & + & b \, \mathcal{V} \, \Delta t \, \mathscr{G}_p(t_n+\alpha \,\Delta t, E^n_q + \,\beta \, \Delta t \, \mathscr{G}_p(t_n,E^{n}_q))+ o((\Delta t)^2)
\end{array}
\end{equation}
where $\Delta t$ is the time step and $t_n$ is the time at step $n$.
For our calculations, we used $a=0$, $b=1$, $\alpha=\frac{1}{2}$ and $\beta=\frac{1}{2}$.

In order to ensure the scheme stability during the time integration, a CFL-like condition must be respected at each time step \citep[chapter 6.3]{1928MatAn.100...32C,Toro_inbook}.
In the absence of magnetic fields and viscosity, this condition is fulfilled as long as the time step is less than:
\begin{equation}\label{eq:cfl condition}
\Delta t_0 = \min\left( N_{CFL} \frac{\mathcal{V}}{\left( ||\vec{V}|| + c_s \right) \max (||d \vec{A}||)} \right)
\end{equation}
where $c_s$ is the sound speed in the center of the cell and $N_{CFL}$ is a constant parameter, which is equaled to ${1}/{2}$ in order to improve stability.
Physically, the quantity $\max\left\{||\vec{V}|| + c_s\right\}$ is the highest wave speed in the whole computational domain which allows interpreting the CFL requirement as a condition for avoiding multiple wave overlap within cell $i$ when the Riemann problem is solved.
Indeed, only two solutions of the Riemann problem can affect cell $i$: the solution coming from left cell $i-1$ and the solution coming from the right cell $i+1$ \citep[chapter 6.1]{Toro_inbook}.

\subsection{Numerical fluxes and exact Riemann solver}\label{subsec:numerical fluxes and exact riemann solver}

From the intrinsic definition of FVM, a numerical solution possesses discontinuities at inter-cells positions, which forces the fluxes at cell boundaries to be computed.
Indeed, flux terms are step functions, which are not continuous from one cell to the other.
At each time step, $t_n$, a correct evaluation of these fluxes is done thanks to the solution of the following initial value problem (IVP):
\begin{equation}\label{Eq:Riemann problem}
\begin{array}{ccl}
\partial_t E_p & = & \mathscr{G}_p(t,E_q) \\
     & \mbox{with} &  E_p(x,t=0)=\left\{
\begin{array}{lcc}
E_{p}(i-1)   &if& x<x\left(i-\frac{1}{2}\right) \\ [6pt]
E_{p}(i+1)   &if& x>x\left((i+\frac{1}{2}\right)
\end{array}\right. 
\end{array}
\end{equation}
This corresponds to a local Riemann problem, the solution of which is executed numerically thanks to an exact Riemann solver which has been taken from \citet[chapter 4]{Toro_inbook}.
Even if $\mathscr{G}_p(t,E_q)$ contains source terms, which does not pose calculation problems, the Riemann solver is more sensitive to the scheme permitting to estimate fluxes.
Indeed, a difficulty is that Euler variables are only defined on cell's centers, but flux terms need to be computed on cell's frontiers.
A workaround is to improve the FVM with a more advanced spatial scheme, which could permit better estimation of fluxes between cells and, thus, obtaining accurate solutions even in the presence of shocks.
Therefore, for this version of \rossbi{}, we use \textbf{(1)} a three-point parabolic interpolation of Euler variables, which allows accurate computation of flux terms at cell's boundaries, combined with \textbf{(2)} a MinMod flux limiter in order to avoid undesired oscillations.

All steps of this section are summarised below (for cell $i$).
First, we perform the \textbf{(1)} Euler variables interpolation at the cell frontiers.
For instance, the left and right Euler variables at frontier $i+1/2$, $E_p^l\left(i+1/2\right)$, and $E_p^r\left(i+1/2\right)$, respectively, are estimated thanks to the parabolic interpolation described in Sect. \ref{subsec:parabolic interpolation}.
Then, \textbf{(2)} fluxes are evaluated in the right and left side of each frontier.
Still for frontier $i+1/2$, we get:
\begin{align}
\mathscr{F}^r_p(i+1/2)=\mathscr{F}_p(E_q^r(i+1/2)) \\
\mathscr{F}^l_p(i+1/2)=\mathscr{F}_p(E_q^l(i+1/2)) 
\end{align}
Finally, we proceed to the \textbf{(3)} Riemann problem resolution.
When fluxes are estimated for the first time at frontiers $i-1/2$ and $i+1/2$, the exact Riemann solver allows computing exactly the Euler variables at frontiers and then reevaluating fluxes:
\begin{equation}
\begin{array}{lll}
E_p^*           & = & \textit{Riemann}(E^l_q, E^r_q) \\
\mathscr{F}_p^* & = & \mathscr{F}_p(E_q^*) 
\end{array}
\end{equation}
where \textit{Riemann} stands for the solver.
We highlight that the choice of the exact Riemann solver is justified by the absence of viscosity in the simulation box.
Indeed, this solver was specially designed for the treatment of Euler's equations.

\subsection{Method improvement thanks to the stationary solution and a parabolic interpolation}\label{subsec:parabolic interpolation}

The equilibrium solution presented in Eq. \ref{Eq: equilibrium solution} could be used to improve the numerical scheme described in Sect. \ref{subsec:numerical fluxes and exact riemann solver}.
First, we rewrite the Euler variables under the form:
\begin{equation}
\begin{array}{ccl}
E_p(i) &=& E_p^0(i) + E_p'(i)
\end{array}
\end{equation}
where $E_p^0$ are the stationary solution Euler variables.
$E_p'$ stands for deviations to the steady axisymmetric flow: for the stationary solution, we get $E_p'=0$.
Thanks to this formulation, the Euler variables at cell $i$ frontiers are:
\begin{equation}
E_p(i\pm{1}/{2}) = E_p^0(i\pm{1}/{2}) + E_p'(i\pm{1}/{2})
\end{equation}
Such formulation has two advantages.
First, the stability of the stationary solution is improved since we use the exact value of the Euler variables at the cell frontiers and subsequently the exact flux terms.
Second, the accuracy of flux terms is improved thanks to a more accurate interpolation of Euler's variables described in \ref{subsec:numerical fluxes and exact riemann solver}.
Within this maneuver a parabolic interpolation on $E_p'$ permits obtaining the left and right Euler's variables at cell's frontiers:
\begin{equation}
\begin{array}{ccl}
E_p^l(i+1/2) &=& E_p^0(i+1/2) + a \left[ x(i+1/2)-x(i)\right]^2 \\[4pt]
             & & + b \left[ x(i+1/2)-x(i)\right] + E_p'(i)      \\[6pt]
E_p^r(i-1/2) &=& E_p^0(i-1/2) + a \left[ x(i-1/2)-x(i)\right]^2 \\[4pt]
             & & + b \left[ x(i-1/2)-x(i)\right] + E_p'(i) 
\end{array}
\end{equation}
where:
\begin{equation}
\begin{array}{ccl}
a &=& \left[\nabla E_p'(i+1) - \nabla E_p'(i) \right] \Big/ \left[ x(i+1)-x(i-1) \right]  \\[6pt]
b &=& \left[\nabla E_p'(i+1)\,(x(i)-x(i-1)) + \nabla E_p'(i)\,(x(i+1)-x(i)) \right]       \\[6pt]
  & & \qquad \qquad \Big/ \left[ x(i+1)-x(i-1) \right]
\end{array}
\end{equation}
In order to avoid undesired oscillations, we use a MinMod flux limiter which, when necessary, reduces the curvature of the interpolation function \citep[Appendix]{survi2013}.

\subsection{Balanced scheme}\label{subsec:balanced scheme}

\begin{figure}
\centering
\includegraphics[trim = 0.0cm 0cm 0cm 0cm, clip, width=0.9\hsize]{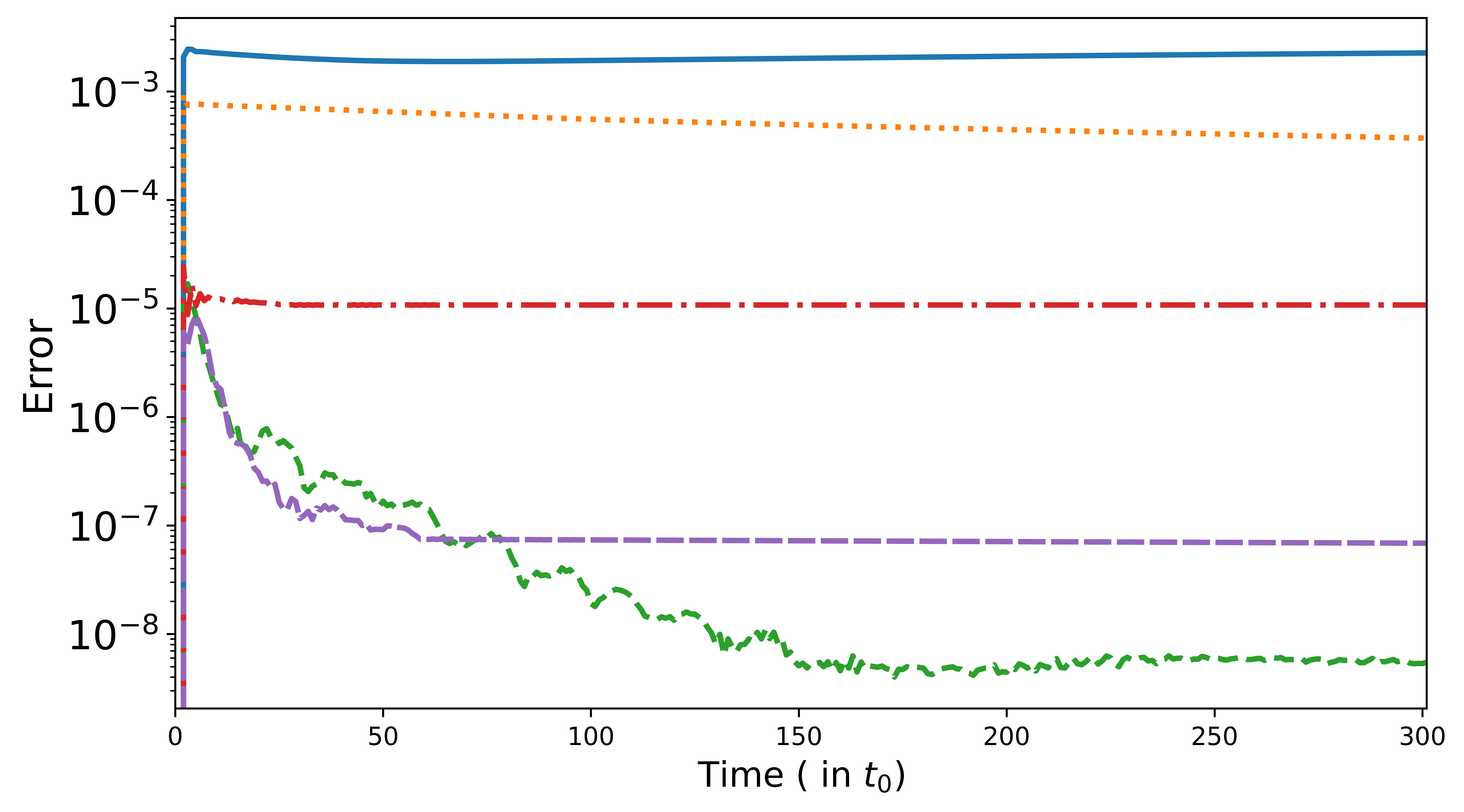}
\includegraphics[trim = 0.0cm 0cm 0cm 0cm, clip, width=0.9\hsize]{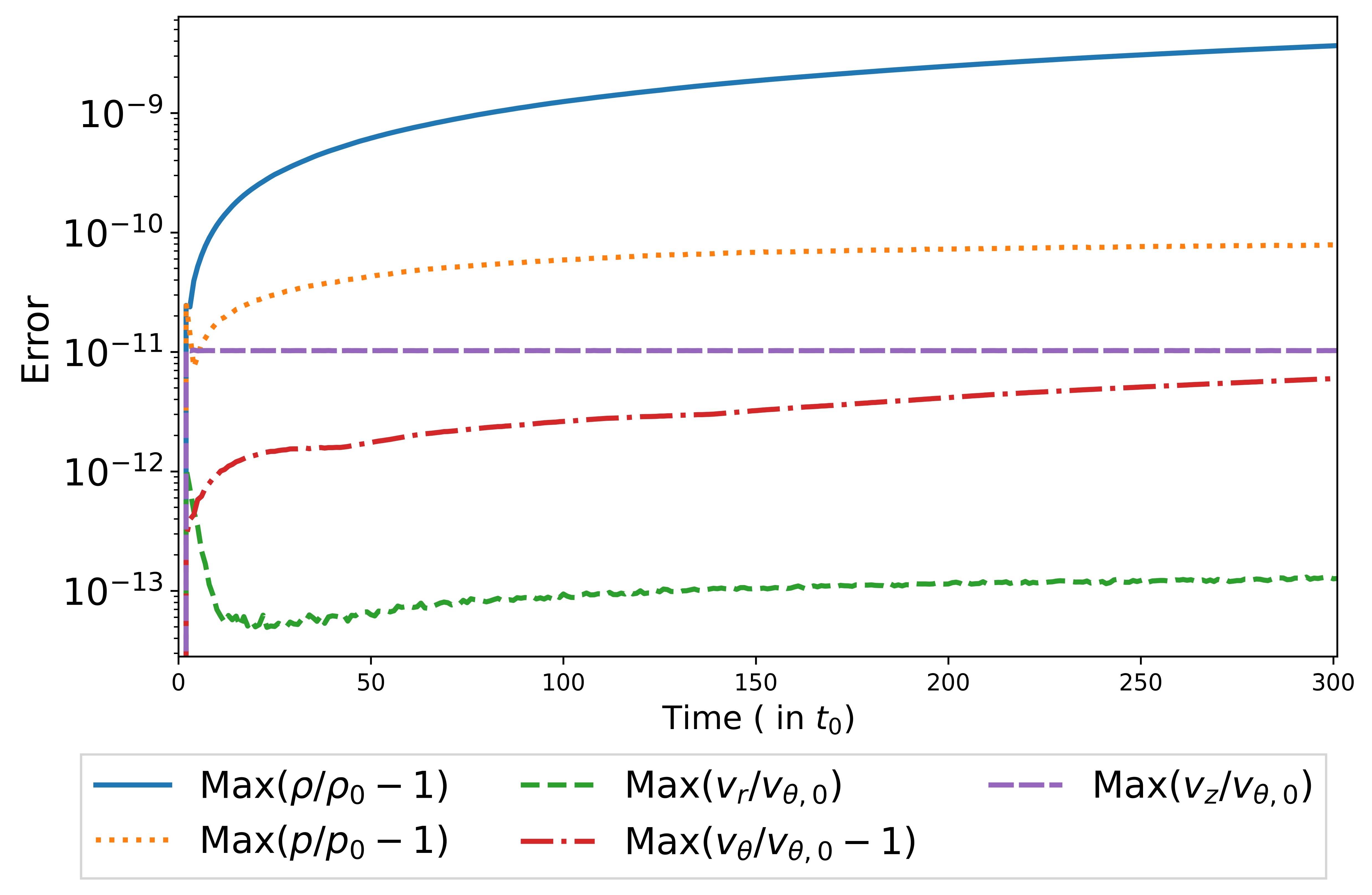}
\caption{\textbf{Errors for the balanced and unbalanced numerical schemes.} \\
\emph{Top:} Unbalanced \\
\emph{Bottom:} Balanced } 
\label{fig:error balanced scheme}
\end{figure}

One sensitive issue in PPD is that the stationary solution is not in a numerical equilibrium.
Indeed, the scheme induces residual terms, which affect the stationary solution during the time evolution steps.
For instance, the evolution of the radial velocity, written in terms of Euler variables is:
\begin{equation}\label{Eq:balanced scheme}
\mathcal{V}_{i,j,k}\partial_t E_1 = 
\mathcal{V}_{i,j,k} \frac{P_0(i,k)}{r(i,k)} + 
\mathcal{V}_{i,j,k} \left.\frac{\partial P_0}{\partial r}\right|_{(i,k)}  -
dz \, d\theta \,\left[r P_0\right]^{(i+\frac{1}{2},k)}_{(i-\frac{1}{2},k)}
\end{equation}
where the right hand side stems from the computation of flux and source terms (see Eq. \ref{Eq:compact form} and Appendix \ref{app:conservative form variables}).
For an ideal infinite mesh resolution, the r.h.s term of above equation vanishes, but in practise it is not the case.
Indeed, this term depends on the accuracy of $\left[r P\right]^{(i+\frac{1}{2},k)}_{(i-\frac{1}{2},k)}$ computation, which introduces a small residual term affecting the stationary solution after few tens of orbits.

At 2D, \citet[Chapter III.2.1]{survi2013} corrected this problem using an interpolation of Euler variables and performing the exact computation of the stationary solution source terms.
However, for the current 3D code, this is a more difficult task since the same method would imply the analytic integration of source terms in the radial and vertical directions.
For above reason, we adopted another technique for the 3D code proposed by Hervé Guillard, which consists on performing an \emph{empty run} which allows to compute the r.h.s residual term for the equilibrium solution.
This residual term is then subtracted at each time step from the right hand side of Eq. \ref{Eq:balanced scheme}.
Such strategy was also adopted by \citet[Sect. 4.4]{2010A&A...516A..31M}.
Other sophisticated techniques exist, but they do not show significant improvements while the present one has the advantage to be easy and rapid to implement.
In Fig. \ref{fig:error balanced scheme}, we report the error in different physical quantities for the equilibrium solution explored in Sect. \ref{subsec:equilibrium solution}, depending on whether the numerical scheme is balanced or not. 
The effectiveness of the exposed method is confirmed: the errors of the unbalanced scheme are 5 orders of magnitude higher than the balanced scheme at least.

\subsection{Aerodynamic forces and evolution of the solid phase}\label{subsec:dust treatment}

In PPDs, particles interact with the gas mainly through friction forces that look like a headwind depending on the size of the particles and the gas free mean path.
From the kinetic theory of gases, the mean-free-path of a gas molecule, $\lambda$, is defined as the length travelled between two successive collisions:
\begin{equation}
\displaystyle\lambda= \frac{\mu \ m_u }{\sigma_{eff}} \frac{1}{\rho}
\end{equation}
where $m_u$, $\mu$, and $\sigma_{eff}$ are the atomic mass, the number of atoms of the gas, and the molecule cross-section, respectively.
The two regimes describing drag forces are the Epstein and Stokes regime.
The former describes particles with sizes smaller than the gas mean-free-path interacting with the gas through molecular collisions and is equivalent to a ram pressure force.
The latter deals with particles with a size being greater than the gas mean-free-path and experiencing a viscous fluid friction.
For ideal spherical particles, the laws of friction expressions are well known \citep{1977MNRAS.180...57W,2010EAS....41..187Y};
the norm of the volume drag force is $||\vec{f}_{drag}||=\rho_d \, { ||\Delta \vec{v}||}/{t}_{stop}$ with:
\begin{equation}
t_{stop} = 
\left\{
\begin{array}{lllll}
\displaystyle t_{stop}^E = \frac{\rho_s \, r_p}{\rho_g \, c_s}
& \mbox{if} &
r_{p} \leq \displaystyle\frac{9 \lambda}{4} & \mbox{(Epstein regime)} \\ [10pt]
\displaystyle t_{stop}^S = \frac{4}{9} \frac{r_p}{\lambda} t_{stop}^E
& \mbox{if} & 
r_{p} > \displaystyle\frac{9 \lambda}{4} & \mbox{(Stokes regime)}
\end{array}\right.
\end{equation}
where $r_p$, $\rho_s$, and $\Delta \vec{v}$ are the radius, the internal density of particles, and particles velocity relative to the gas, respectively.

Solid material is considered as a pressureless fluid interacting with gas through the aerodynamic forces introduced in the above paragraph.
Therefore the equations governing dust dynamics are similar to the ones of gas except that the pressure term vanishes, which implies that there is no energy equation to be solved for this phase.
The absence of pressure implies that the Riemann solver should be adapted to solid particles, but essentially it is almost the same as for gas.

\subsection{Self-gravity and indirect-gravity}\label{subsec:SG and indirect}

The current version of \rossbi{} supports SG for 2D simulations and accounts for the correction and generalisation of the smoothing length method to bi-fluids proposed by \citet{rendonbarge_2023}.
On the other hand, the 3D version lacks SG calculation.
Indeed, this is a long-term task requiring care in the theoretical and programming stages.
Yet, we already thought about it and we provide a discussion on different possibilities for future versions in Sect. \ref{subsec:3D selfgravity}.

The presence of a disc asymmetry can lead to an offset of the mass barycenter \citep{2016MNRAS.458.3918Z,2017A&A...601A..24R}. 
We have taken this effect into account through the gradient of the indirect gravity potential, which can be read in cylindrical coordinates:
\begin{eqnarray}
\Vec{\nabla} \Tilde{\Phi}_{ind} & = & X_{ref} \, 
\left(
\begin{matrix} 
A_{disc} \cos{\theta}    + B_{disc} \sin{\theta}  \\ 
- A_{disc} \sin{\theta}  + B_{disc} \cos{\theta}  \\
0
\end{matrix}\right)
\end{eqnarray}
where $A_{disc}=\int\limits_{disc} 
                \frac{\Tilde{r}' \cos{(\theta')} }{\left(\Tilde{r}'^2+\Tilde{z}'^2\right)^{3/2}} \, d\Tilde{m}'$
and $B_{disc}=\int\limits_{disc} 
              \frac{\Tilde{r}' \sin{(\theta')}}{\left(\Tilde{r}'^2+\Tilde{z}'^2\right)^{3/2}} \, d\Tilde{m}'$.
We assumed that the density vertical stratification is symmetric with respect to the $z=0$ plane that permits the vertical component of the indirect gravity to be cancelled. 

\subsection{Numerical resolution summary}

\begin{table}[]
\centering
\begin{tabular}{l c c}
\hline
Version                            & 2D      & 3D     \\
\hline
Box                                & yes     & yes    \\
Parallelism in all directions      & yes     & yes    \\
Dust drag                          & yes     & yes    \\
Dust back-reaction                 & yes     & yes    \\
SG                       & yes     & no     \\
Indirect-Gravity                   & yes     & yes    \\
Viscosity                          & no      & no     \\
Magnetic fields                    & no      & no     \\
\hline
\end{tabular}
\caption{Options available in the 2D and 3D version}
\label{tab:options}
\end{table}

We presented the new bi-fluid \rossbi{} code, which is the natural extension of \emph{RoSSBi} to 3D.
This 3D version is based on a FVM scheme, is third order in space, second order in time, stable, balanced, and accounts for inherent boundary instabilities, which could arise in PPDs simulations. 
The current version of \rossbi{} supports the options gathered in Table \ref{tab:options}.
In the next section, we show how we succeeded in making a fast and scalable 2D and 3D code well-suitable for HPC.

\section{Improving performances on parallel architectures}\label{sec:improving performances on parallel architectures}

Parallelism in this code is implemented using an MPI programming model \citep{mpi40}. 
It is a standardised protocol that provides a communication among processes running on parallel computers, typically consisting of many nodes with distributed memory model.

\subsection{Parallel decomposition of the regular grid}\label{subsec:domain decomposition}

\begin{figure}
\centering
\includegraphics[width=\hsize]{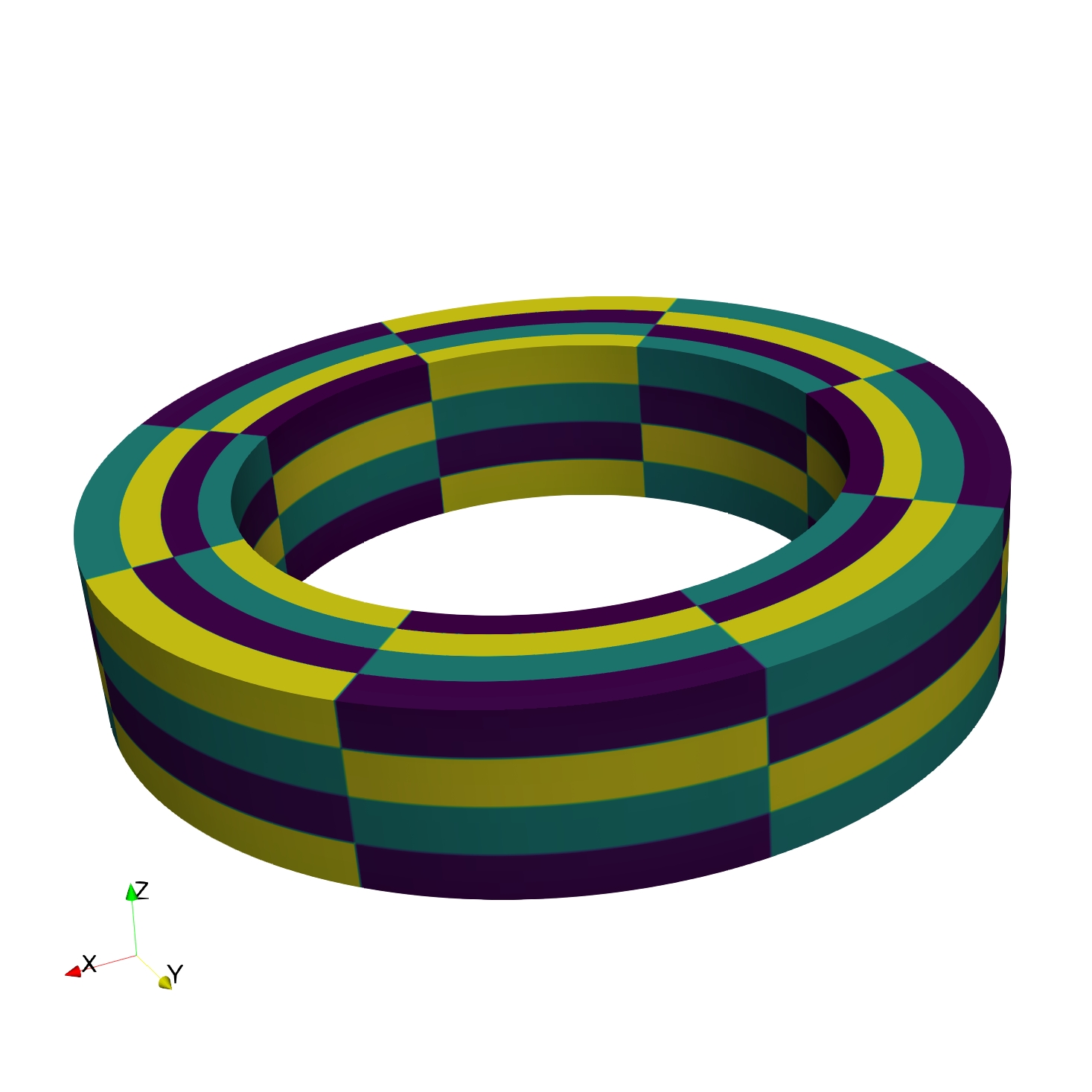}
\caption{\textbf{3D domain decomposition on 128 processors.}\\
The subdomains of same colour are not treated by the same processor
(It permits only to differentiate subdomains from each other).}
\label{fig:parallel domain}
\end{figure}

\begin{table}[]
\centering
\begin{tabular}{lll}
\hline
                                           & \multicolumn{1}{c}{\textbf{1D}}           
                                                                   & \multicolumn{1}{c}{\textbf{2D}} \\
\hline\rule{-2pt}{1.2\normalbaselineskip}
$\mathcal{V}_{comp}$                       & $\propto {N^2}/{P}$   &  $\propto {N^2}/{P} $      \\
$\mathcal{S}_{comm}$                       & $\propto N$           &  $\propto  N/\sqrt{P}$     \\
$CC=\mathcal{S}_{comm}/\mathcal{V}_{comp}$ & $\propto {P}/{N}$     &  $\propto {\sqrt{P}}/{N}$  \\
\hline
\end{tabular}
\caption{Parallel domain decomposition (1D and 2D) for a 2D computational box.}
\label{tab:domain decomposition}
\end{table}

Domain decomposition is one key to reach scalability and high-performance at high resolutions.
As an instructive example, let us consider a 2D square computational box containing $N$ cells in each direction where the computational effort (workload) is shared among $P$ processors.
In a crude approach, the workload computed by a processor is proportional to the volume of data the assigned cells contain ($\mathcal{V}_{comp}$) while communication (the data to be transferred among processors) is proportional to their surface ($\mathcal{S}_{comm}$).
One may use the communication-to-computation ratio, $CC=\mathcal{S}_{comm}/\mathcal{V}_{comp}$\footnote{In numerical code metrics, we often use granularity defined by $G=T_{comp}/T_{comm}$ where $T_{comp}$ and $T_{comm}$ are the measured computation and communication times, respectively.}, which helps to roughly estimate the bandwidth requirements.
Since the ratio between the available bandwidth and the computing power of the modern parallel HPC systems is typically low, it is also important to keep the $CC$ ratio low during the algorithm design.
For distributing the $P$ processors in this regular grid, we consider 1D and 2D domain decompositions the features of which are gathered in Table \ref{tab:domain decomposition}.
The $CC$ ratio is lower in the 2D decomposition compared to the 1D decomposition since it grows more slowly with $P$.
Therefore, communication costs are smaller in the 2D case and more suitable for architectures with low bandwidth.
For the reasons above, it was decided to \textbf{(1)} change the 1D grid decomposition to a 2D grid decomposition in the 2D code and \textbf{(2)} use a 3D grid decomposition in the 3D code.
Fig. \ref{fig:parallel domain} illustrates how the computational grid could be decomposed and distributed among 128 processors.

\subsection{Performance analysis and optimisations}\label{subsec:MPI optimisation}

\begin{table}[]
\centering
\begin{tabular}{|l|l|l|}
\hline
Version                 & original & updated \\
\hline
Number of processes (total)       & 576     & 576     \\
Number of processes (X, Y, Z)     & (8, 36, 2)     
                                            & (8, 18, 4)     \\
Elapsed time (sec)      & 7.59    & 1.56    \\
Efficiency              & 0.18    & 0.95    \\
Speedup                 & 2.85    & 15.13   \\
Average IPC             & 1.05    & 0.92    \\
Average frequency (GHz) & 2.52    & 3.25    \\
\hline
\end{tabular}
\caption{Metrics overview of the initial and updated version of the code. The efficiency and speedup are relative to the corresponding base runs (36 processes).}
\label{tab:my-table}
\end{table}

Many bottlenecks intrinsically related to the structure of the code may suppress the potential of parallelisation.
Therefore, we took advantage of the offer by Performance Optimisation and Productivity CoE (POP) that helps users and developers to optimise parallel software and understand performance issues.
The initial performance assessment of the \rossbi{} code was requested to identify areas of improvement in large resolution simulations. 
The assessment was conducted using Barbora cluster at IT4Innovations, National Supercomputing Center at VSB – Technical University of Ostrava.
The cluster consists of 192 powerful x86-64 compute nodes, each equipped with 2x 18-core Intel Cascade Lake 6240 at 2.6 GHz processors and 192 GB of DDR4 memory, interconnected by high-speed InfiniBand HDR100 network.
As the vast majority of the current supercomputers, Barbora is operated by a Linux OS, specifically Red Hat Enterprise Linux 7.9. 
To build and install the \rossbi{} code, the GCC 9.3.0 toolchain together with OpenMPI 4.0.3 was used. 
In addition, the BSC performance toolset including Extrae 3.8.3, Paraver 4.9.0, Dimemas 5.4.2, and Basic Analysis 0.3.9 was employed to collect and analyse performance data.

\begin{figure}
\centering
\includegraphics[width=\hsize]{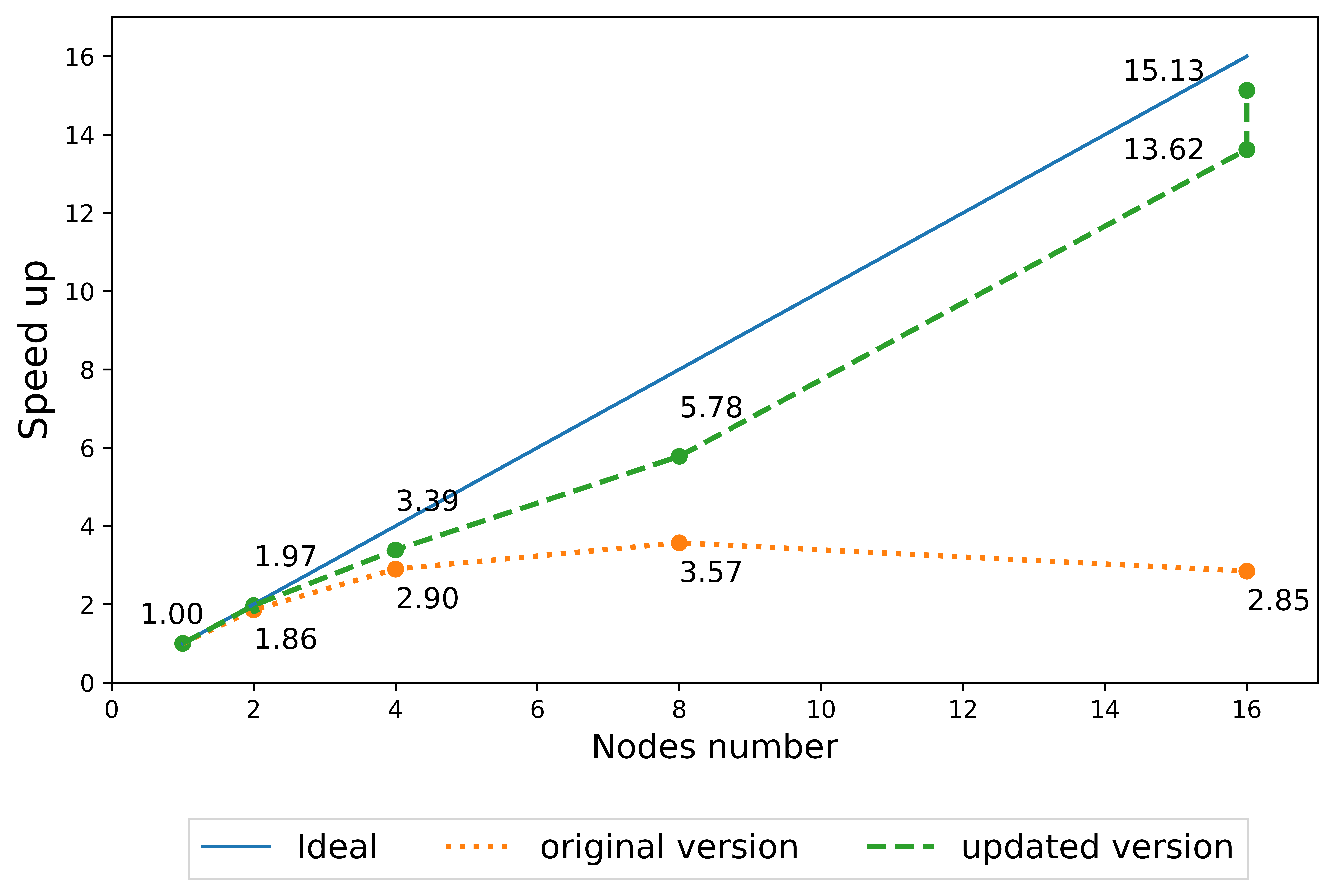}
\caption{\textbf{Scalability of the original and updated version of the \rossbi{} code.}\\
\textbf{Outline: Column Gaussian vortex evolution.} \\
\textbf{Barbora:} 2 Intel Cascade Lake, 18 cores, 2.6 GHz (namely 36 cores/node). Infiniband HDR 200 Gb/s}
\label{fig:scaling_orig}
\end{figure}

The scaling of the \emph{RoSSBi} code was evaluated on 1 – 16 compute nodes, i.e 36 – 576 cores, respectively. 
During the initial stage of the analysis, it turned out that full-length simulations comprising millions of time steps produce an unmanageable amount of collected performance data. 
Using high-level techniques for detecting the computation structure, it was proved that the performance characteristics do not evolve in time significantly.
Therefore, further analysis was done with the simulation duration being reduced to the order of low hundreds of time steps.
The detailed computation structure analysis showed that the code had issues with computation granularity that were indicated by the very short computation bursts – the sequences of user-code instructions outside the MPI library runtime.
Most of the bursts were shorter than 500 $\mu$s.
The scalability tests showed rapid speedup degradation on more than 2 nodes as depicted in Fig. \ref{fig:scaling_orig} (original version). 

The following investigation revealed significant growth of the communication phase that consisted of a sequence of loops with many short, mostly collective, MPI calls. 
Moreover, the loops overhead also caused the increased number of instructions issued. 
Each MPI call or user instruction causes an additional latency that increases the total runtime and effectively degrades the scalability.
To narrow this limiting factor, the code was refactored in the way of fusing the loops and collapsing the thousands of the small MPI messages transferred in each time step into only tens of larger messages.
This enhancement significantly improved efficiency of the inter-node communication through the network and also reduced the instruction overhead.
Another important optimisation was improving the insufficient parallelisation of 3D domains in all directions (see Sect. \ref{subsec:domain decomposition}).
Indeed, the original version of \rossbi{} was only parallelised in the azimuthal direction while the updated version used a 3D domain decomposition.

The follow-on performance assessment demonstrated that the applied optimisations significantly improved the scaling as shown in Fig. \ref{fig:scaling_orig} (updated version). 
During the analysis, we found that also the MPI process configuration and domain decomposition, as  explained in Sect. \ref{subsec:domain decomposition}, has a non-negligible impact on performance. 
For example, we obtained the lower speedup value (13.62) of the updated version in Fig. \ref{fig:scaling_orig} with the 8x36x2 configuration of processes, while the higher value (15.13) using the 8x18x4 configuration.
Table \ref{tab:my-table} reports up to 4.9x speedup on the same largest test case.
Also increased average CPU frequency was observed.
For future scaling to even larger test cases, some optimisation opportunities in the communication pattern were also identified. 
The single-core performance might be enhanced by analysing and optimising the vectorisation opportunities.

We also compared the performances of \rossbi{} when compiled with $\{$OpenMPI library + GCC compiler$\}$ and with $\{$IntelMPI library + ifort compiler $\}$ for a standard test on 64 processors in CeSAM.
The test reports a 1.86 speedup when compiled with Intel compilers that we recommend for Intel architectures.
This is particularly interesting since Intel\textsuperscript{\tiny\textregistered} compilers are free to use now.

\section{Test cases}\label{sec:test cases}

This section is devoted to retrieve common results on PPDs literature thanks to the \rossbi{} code.
For the sake of clarity in this whole section, $t_0$ and $H_g$ stand for the orbital period and the pressure scale height, respectively, at the center of the simulation box ($r_0$).
We start assessing our numerical method with the 2D explosion test.

\subsection{Cylindrical explosion test}

\begin{figure}
\centering
\includegraphics[trim = 0.0cm 0cm 0cm 0cm, clip, width=\hsize]{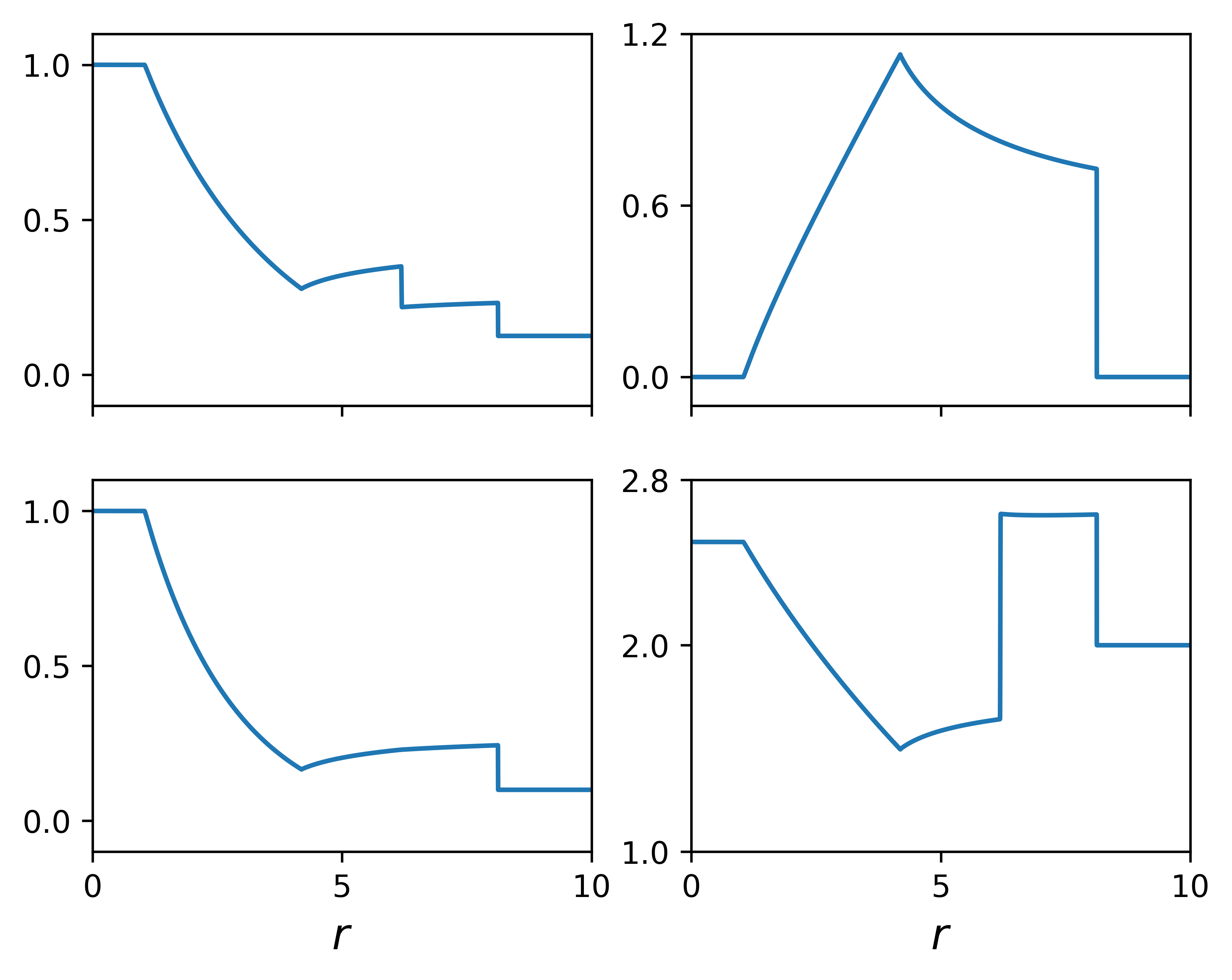}

\caption{\textbf{Cylindrical explosion test at $t=2.5$.} \\
\begin{tabular}{ll}
\emph{Top left:} density      & \emph{Top right:} radial velocity   \\
\emph{Bottom left:} pressure  & \emph{Bottom right:} internal energy  
\end{tabular}
}
\label{fig:shock test}
\end{figure}

The explosion test is a 2D analogue of the sod shock-tube test which permits both the accuracy of a numerical scheme and its ability to keep the cylindrical symmetry of the numerical solution to be assessed \citep[chap. 17]{Toro_inbook}.
This test consists of a 2D Riemann problem in which, as an initial state, a high pressure region and a low pressure region are separated by a diaphragm:
\begin{equation}
(\rho, P, v_r, v_\theta)=
\left\{
\begin{array}{lll}
(1.0, 1.0, 0, 0)       & \mbox{if } &  r \leq 0.4 \\
(0.125, 0.1, 0.0, 0.0) & \mbox{if } &  r > 0.4    
\end{array}\right.
\end{equation}
For this test, we take the ratio of specific heats $\gamma=1.4$, and the number of cells in the radial and azimuthal direction is $(N_r, N_\theta)=(1600, 3200)$.
In Fig. \ref{fig:shock test}, we show this test results at $t=2.5$, which are in good agreement with the reference solution of \citet[Fig. 17.4]{Toro_inbook}.

\subsection{Equilibrium solution}\label{subsec:equilibrium solution}

%
%

We use as initial state the equilibrium solution exhibited in Eq. \ref{Eq: equilibrium solution} and checked that this solution remained stationary during 300 orbits.
The error evolution of all physical variables, shown in Fig. \ref{fig:error balanced scheme}, remained smaller than $10^{-9}$,  which is satisfactory.


\subsection{Rossby wave instability}\label{sec: rossby wave instability}




\begin{figure}
\centering
\includegraphics[trim = 0.0cm 0cm 0cm 0cm, clip, width=0.9\hsize]{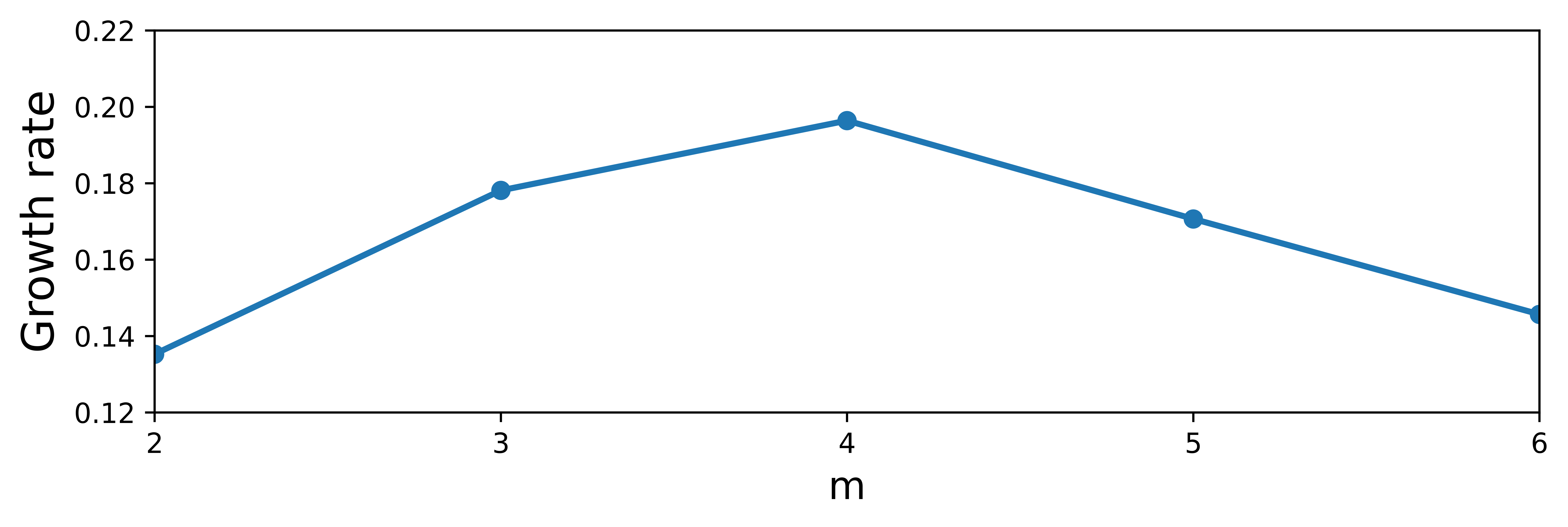}
\caption{\textbf{Rossby wave instability growth rate for each $m$ mode.}} 
\label{fig:RWI growth rate}
\end{figure}

%
\begin{figure*}
\centering

\resizebox{\hsize}{!}
          {\includegraphics[trim = 0.8cm 2cm 1.7cm 9.5cm, clip, width=0.4\hsize]{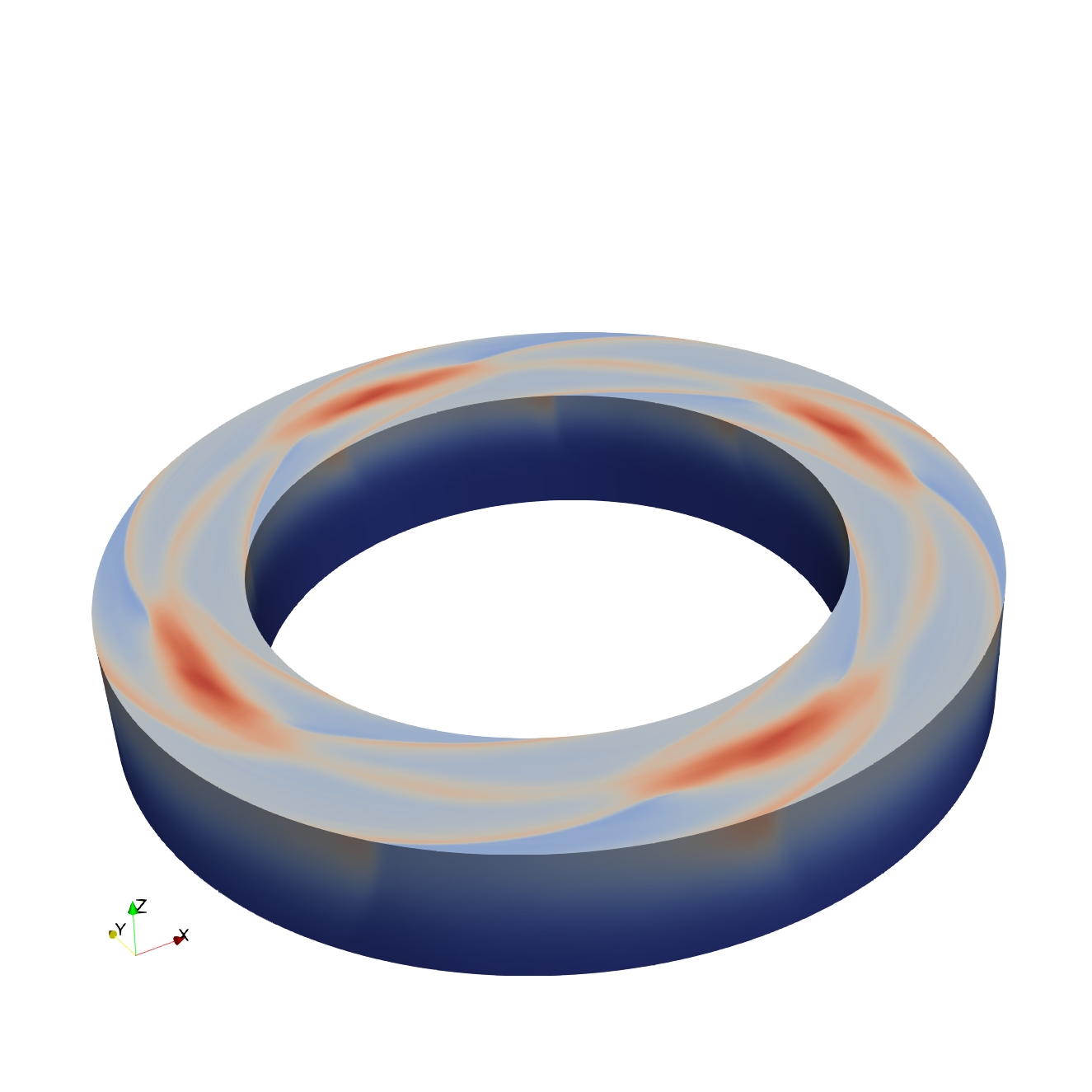}
           \includegraphics[trim = 0.8cm 2cm 1.7cm 9.5cm, clip, width=0.4\hsize]{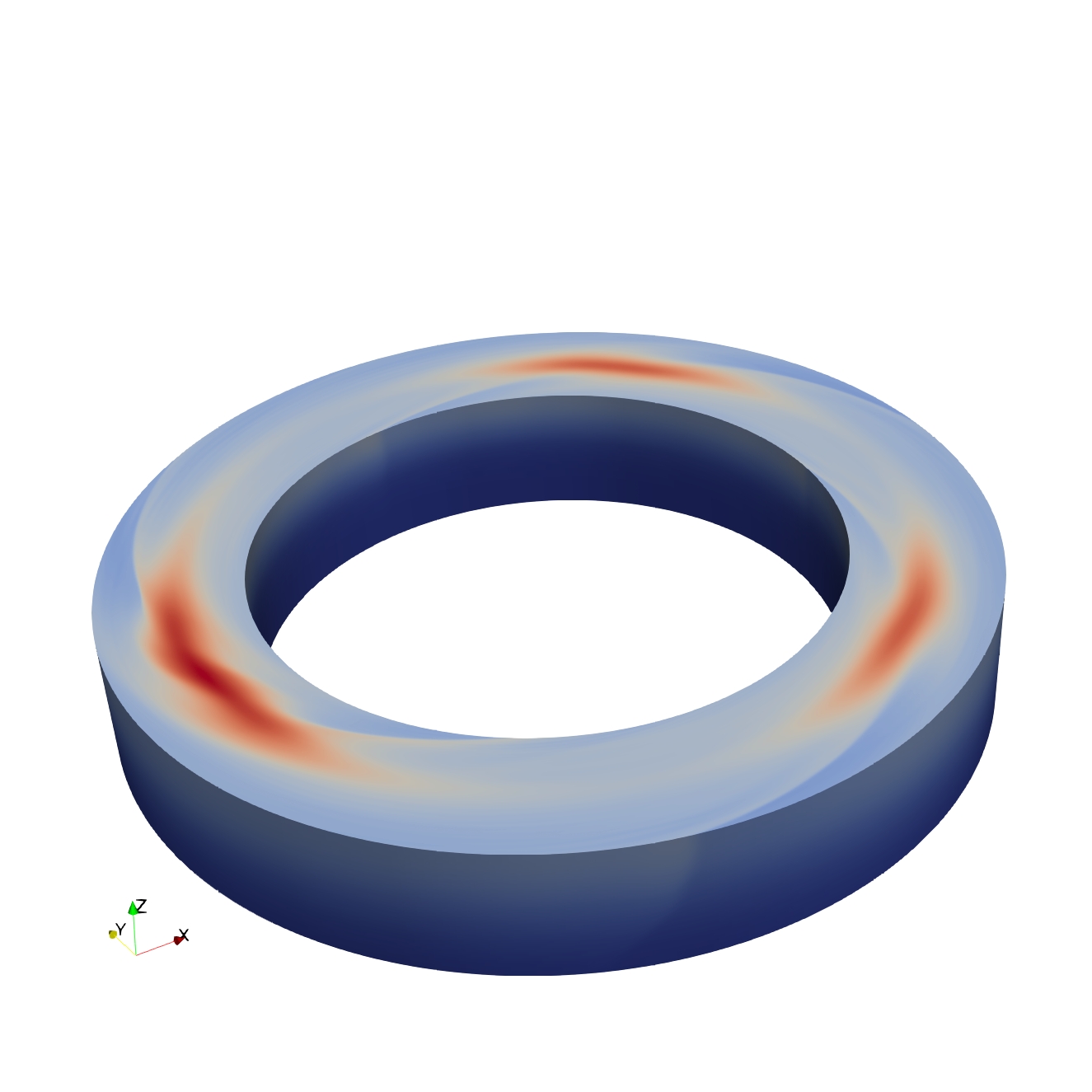}
          }
\resizebox{\hsize}{!}
          {\includegraphics[trim = 0.8cm 2cm 1.7cm 9.5cm, clip, width=0.4\hsize]{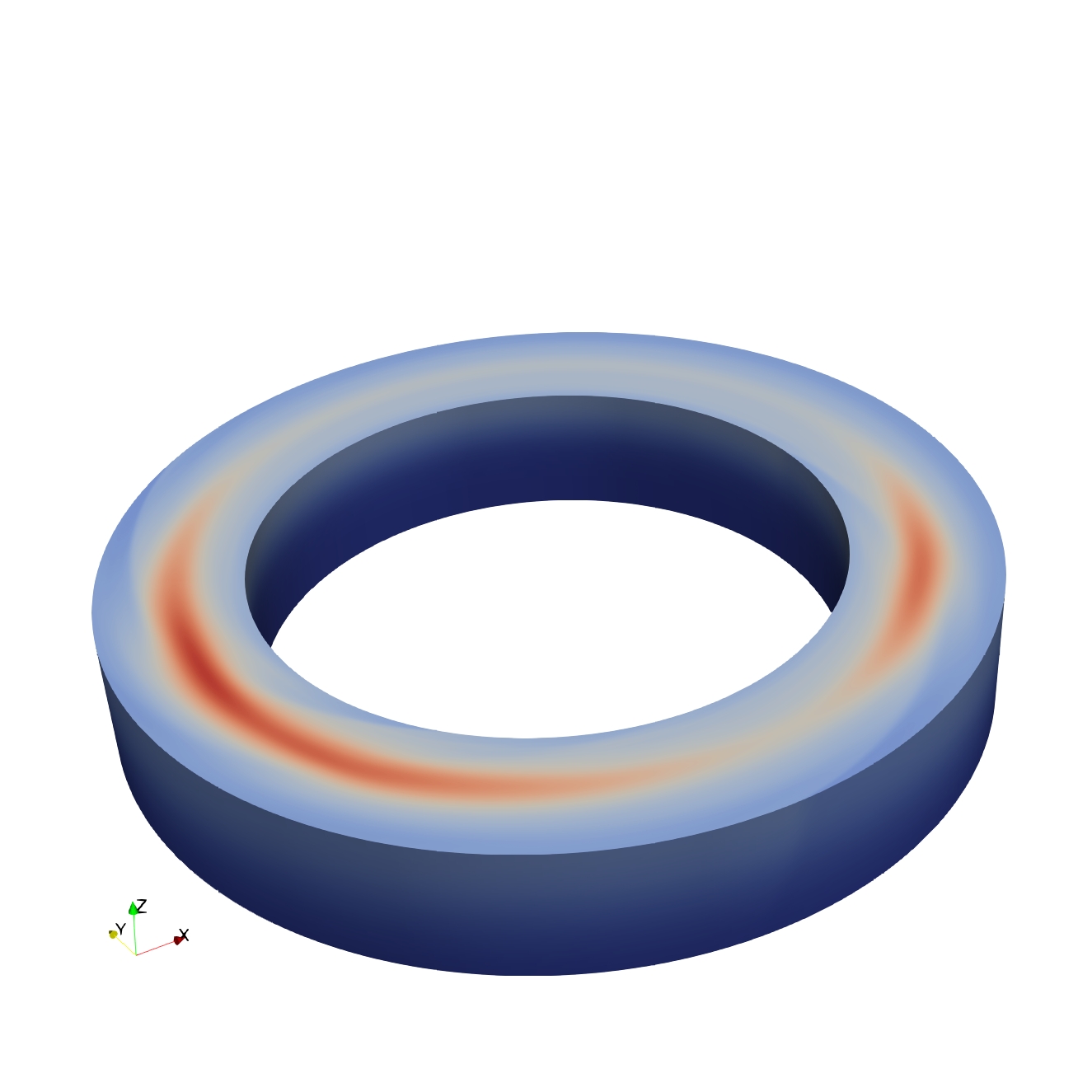}
           \includegraphics[trim = 0.8cm 2cm 1.7cm 9.5cm, clip, width=0.4\hsize]{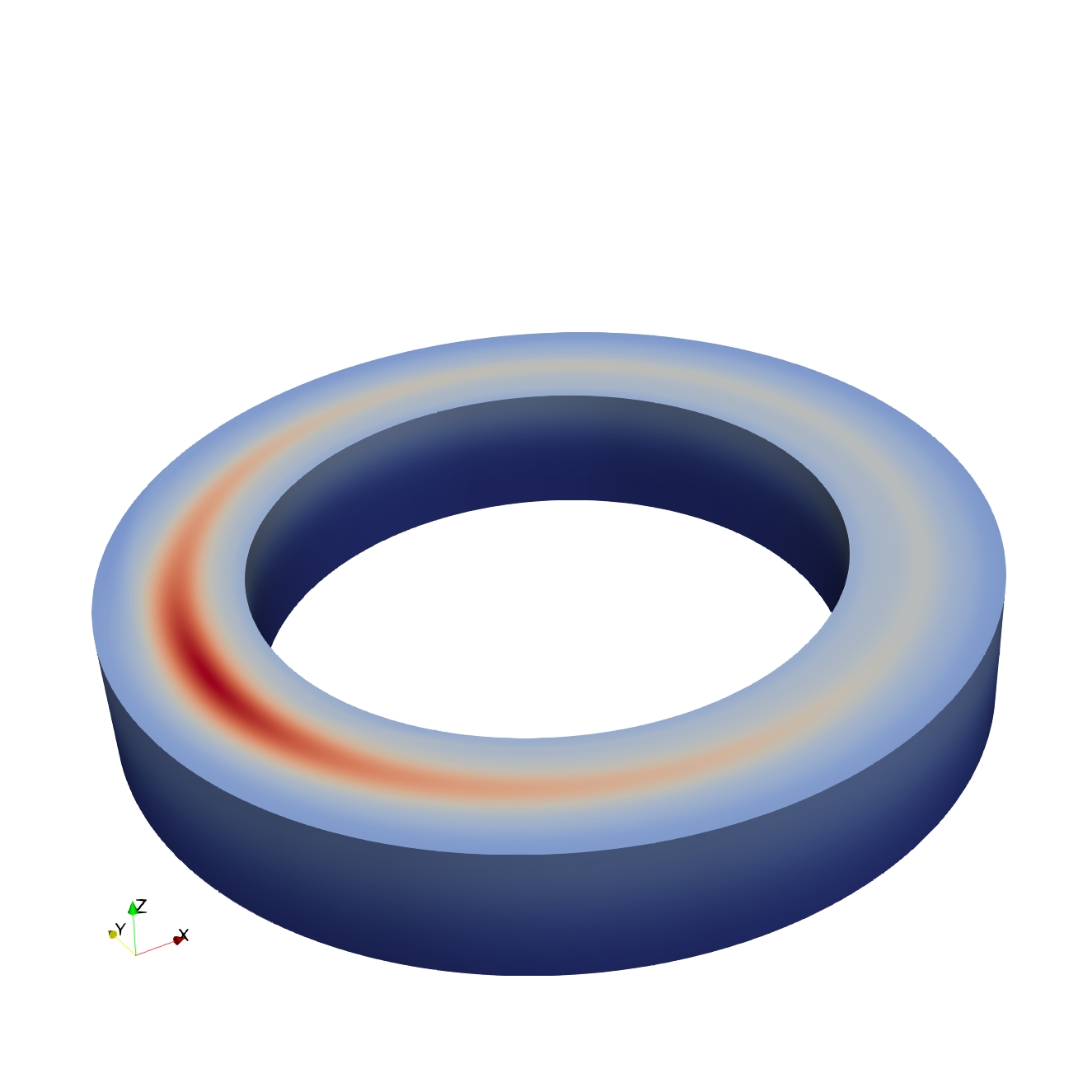}
          }

\caption{\textbf{Rossby wave instability evolution and progressive vortices merging starting from the $m=4$ mode - $\rho_{gas}(r,\theta,z)/\rho_{gas,0}(r,z=0)$} \\
\begin{tabular}{ll}
\emph{Top left:} $m=4$ mode ; $t=11\,t_0$      & \emph{Top right:} $m=3$ mode ; $t=85\,t_0$   \\
\emph{Bottom left:} $m=2$ mode ; $t=210\,t_0$  & \emph{Bottom right:} $m=1$ mode ; $t=330\,t_0$  
\end{tabular}
}
\label{fig:time evolution RWI m=4}
\end{figure*}
%

%
\begin{figure*}
\centering
\resizebox{\hsize}{!}
          {\includegraphics[trim = 5cm 2cm 12.5cm 9.0cm, clip, width=0.4\hsize]{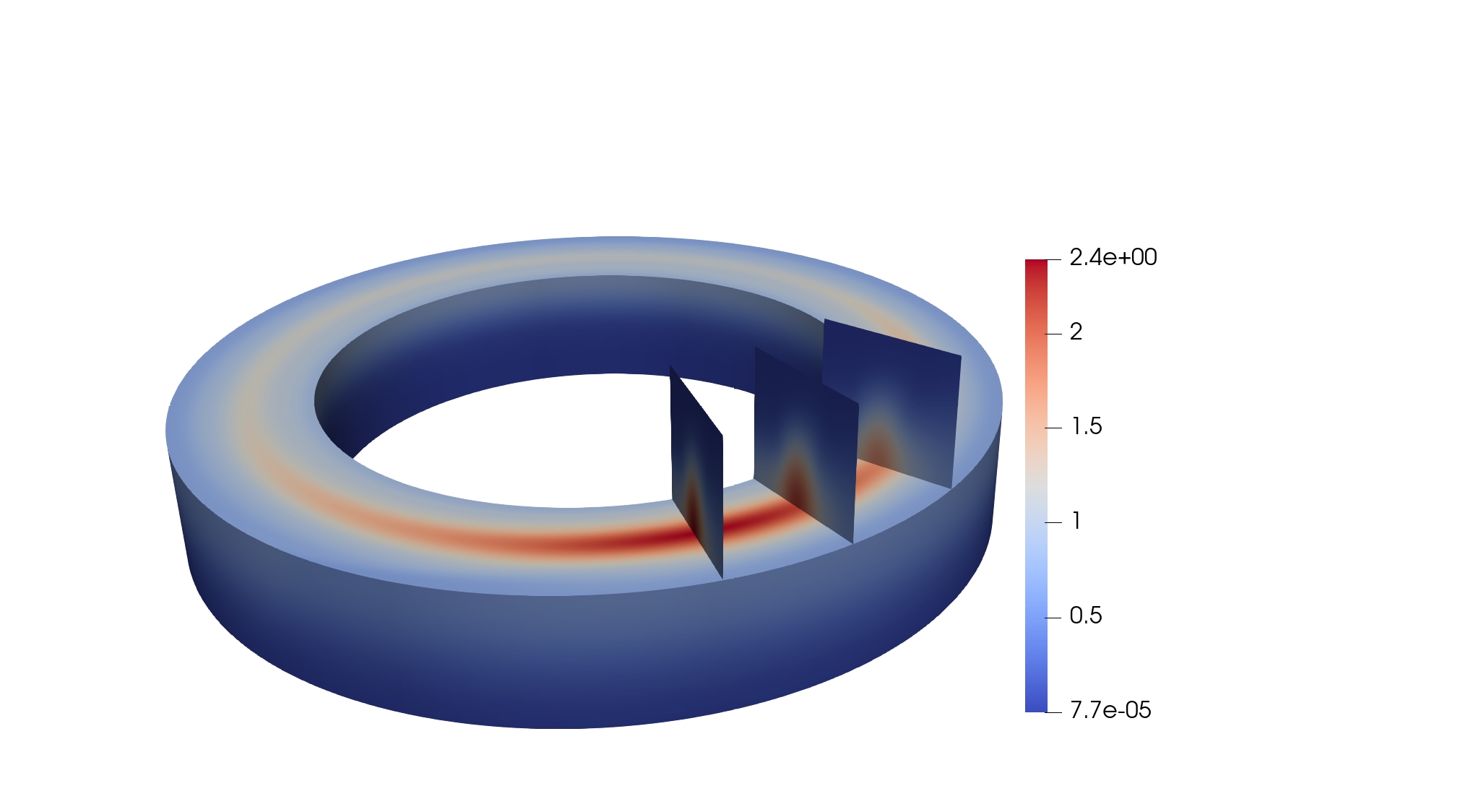}
           \includegraphics[trim = 5cm 2cm 12.5cm 9.0cm, clip, width=0.4\hsize]{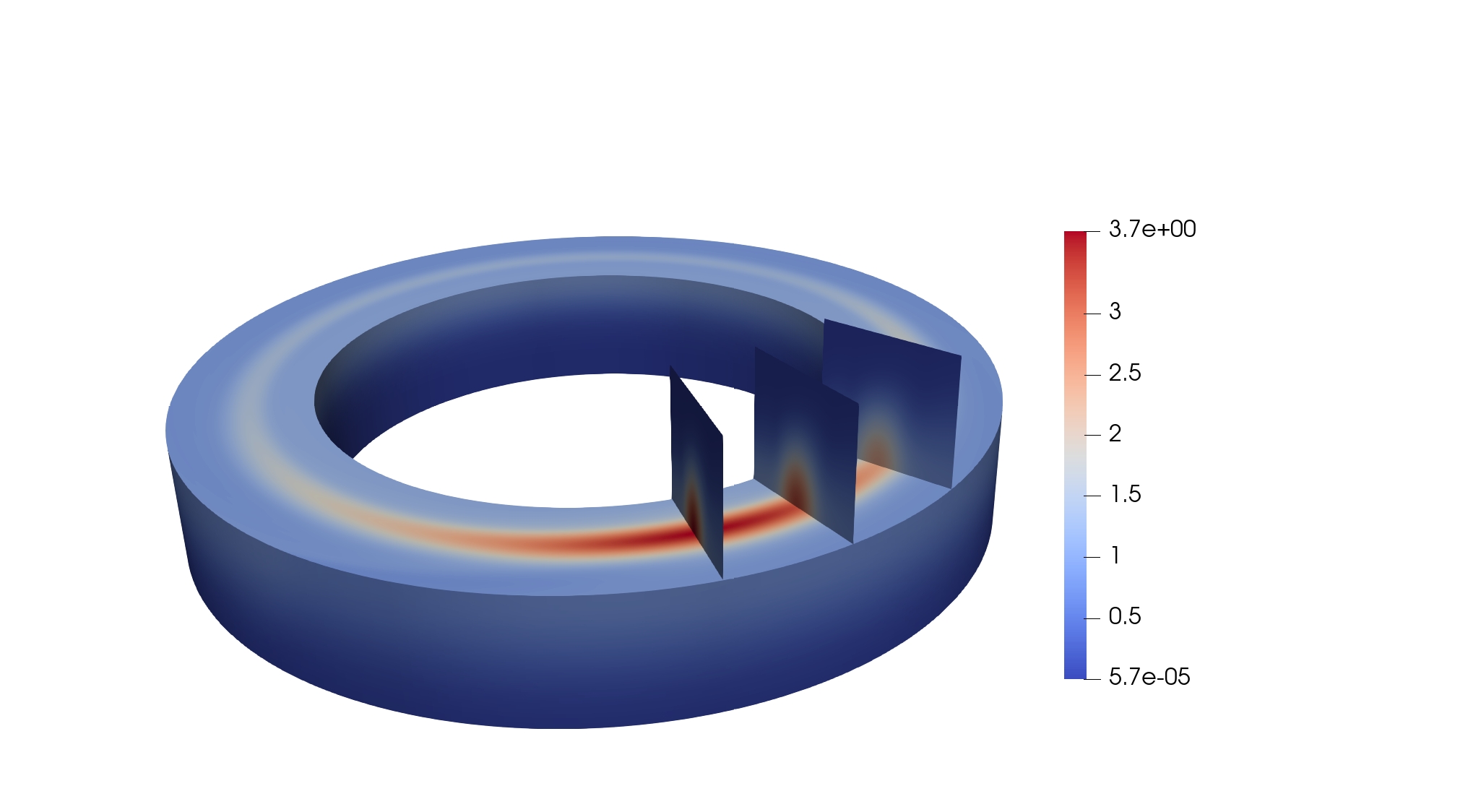}
          }
          
\resizebox{\hsize}{!}
          {\includegraphics[trim = 5cm 2cm 12.5cm 9.0cm, clip, width=0.4\hsize]{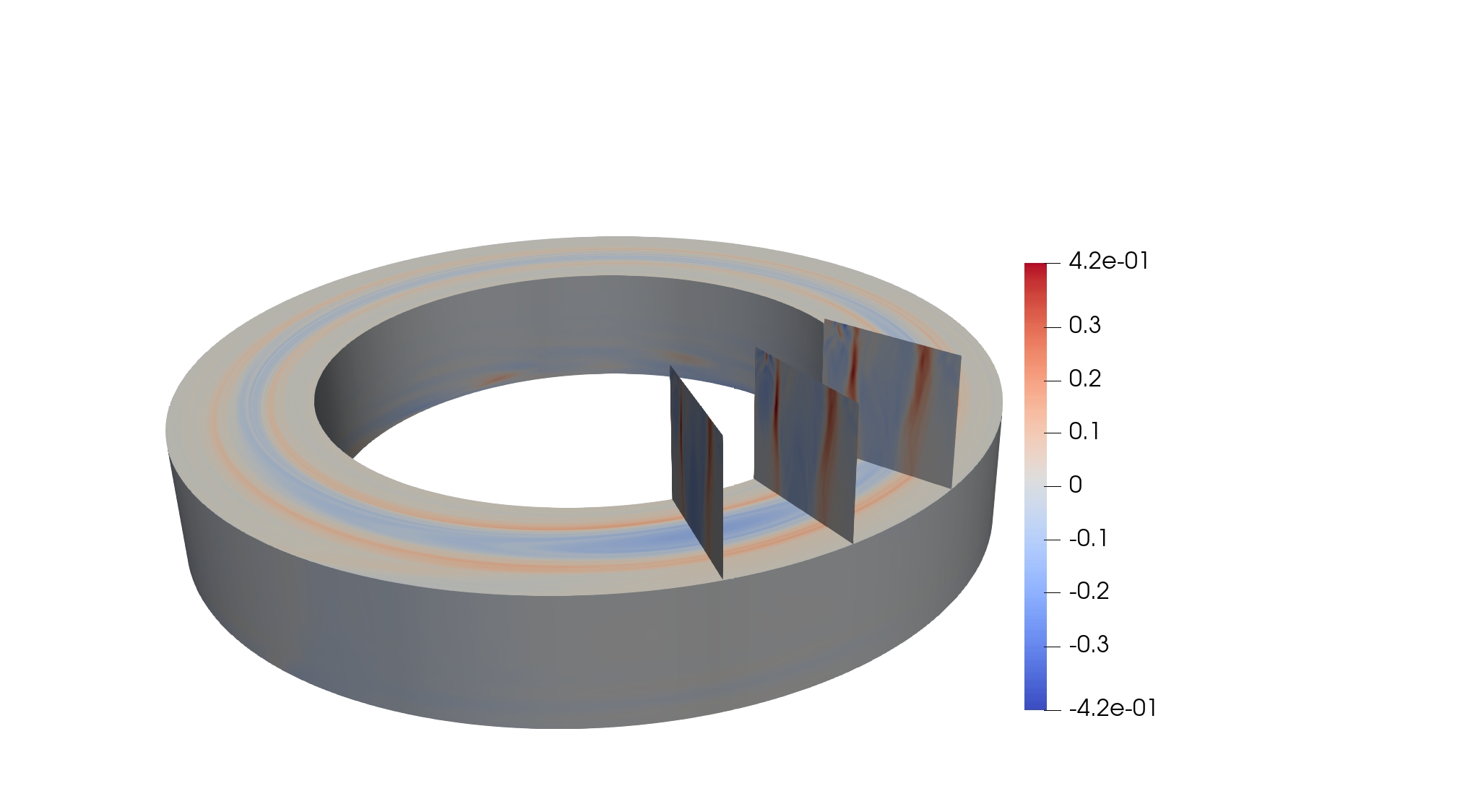}
           \includegraphics[trim = 5cm 2cm 12.5cm 9.0cm, clip, width=0.4\hsize]{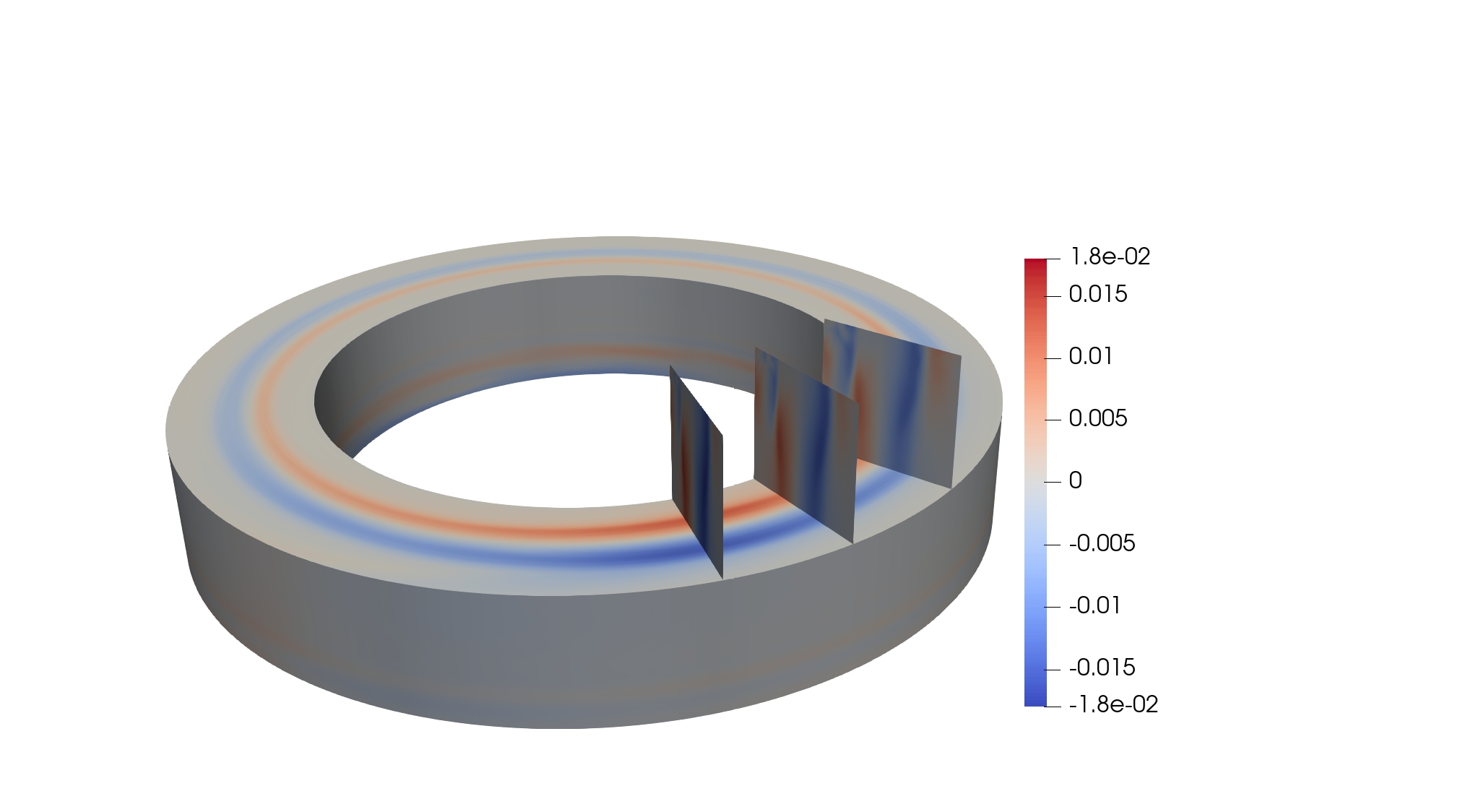}
          }

\caption{\textbf{Mode $m=4$ saturation at the 300$^{th}$ orbit} \\
\begin{tabular}{lll}
\emph{$1^{st}$ row:} & \emph{Left:} Normalised density - $\rho(r,\theta,z)/\rho_{0}(r,z=0)$ 
                     & \emph{Right:} Normalised pressure - $p(r,\theta,z)/p_{0}(r,z=0)$ \\
\emph{$2^{nd}$ row:} & \emph{Left:} Rossby number - $Ro(r,\theta,z)= \omega_z / 2 \Omega_K(r,z)$ 
                     & \emph{Right:} Normalised deviation to azimutal velocity - $v_\theta/v_{\theta,0}-1$ \\
\end{tabular}
\newline
with $\omega_z=\ez \cdot \vec{\nabla} \times \vec{v}'$ and $\vec{v}'=\vec{v}-v_{\theta,0}\et$.}
\label{fig:mode 4 saturation}
\end{figure*}
%

PPDs are subject to instabilities, such as the RWI \citep{1999ApJ...513..805L,2000ApJ...533.1023L}.
Numerical and theoretical work was already considered at 3D for this instability by few authors and in this section, we aim to retrieve \citet{2010A&A...516A..31M,2012A&A...542A...9M} results.
As an initial state, we introduced a radial Gaussian overdensity bump in the stationary solution provided by Eq. \ref{Eq: equilibrium solution}
\begin{equation}
\rho(r,\theta,z)= \rho_0(r,z) \left( 1+\delta \exp{\left[-\left(\frac{r-r_0}{\sqrt{2}\Delta_r}\right)^2\right]} \right)     
\end{equation}
To facilitate the trigger of the unstable modes we introduced a small perturbation in the radial velocity field:
\begin{equation}
v_r(r,\theta,z)=\epsilon \, v_{\theta,0}(r,z) \,
                \sin{\left(m\theta\right)}    \,
                \exp{\left[-\left(\frac{r-r_0}{\sqrt{2}\Delta_r}\right)^2 \right]} 
\end{equation}
where $\epsilon=10^{-4}$ and $m$ is the perturbation mode.
We ran 5 simulation tests for $m=[2,3,4,5,6]$ the outline of which is the same as \citep{2012MNRAS.422.2399M}.
In Fig. \ref{fig:RWI growth rate} we depicted the exponential growth rate obtained for each mode.
Fig. \ref{fig:time evolution RWI m=4} shows the perturbed density evolution in the $z\leq0$ region for the $m=4$ mode up to instability saturation and in Fig. \ref{fig:mode 4 saturation}, we illustrated in detail the vortex generated by the $m=4$ mode and its vertical distribution.
We found that the growth rate values and the curve trend are in agreement with \citet[Fig. 7]{2012MNRAS.422.2399M}.

%


\subsection{Dust trapping by 3D gaseous Vortices}\label{sec: dust trapping by 3D gaseous vortices}

%
\begin{figure*}
\centering
\resizebox{\hsize}{!}
          {\includegraphics[trim = 0.8cm 0cm 12cm 1.5cm, clip, width=0.35\hsize]{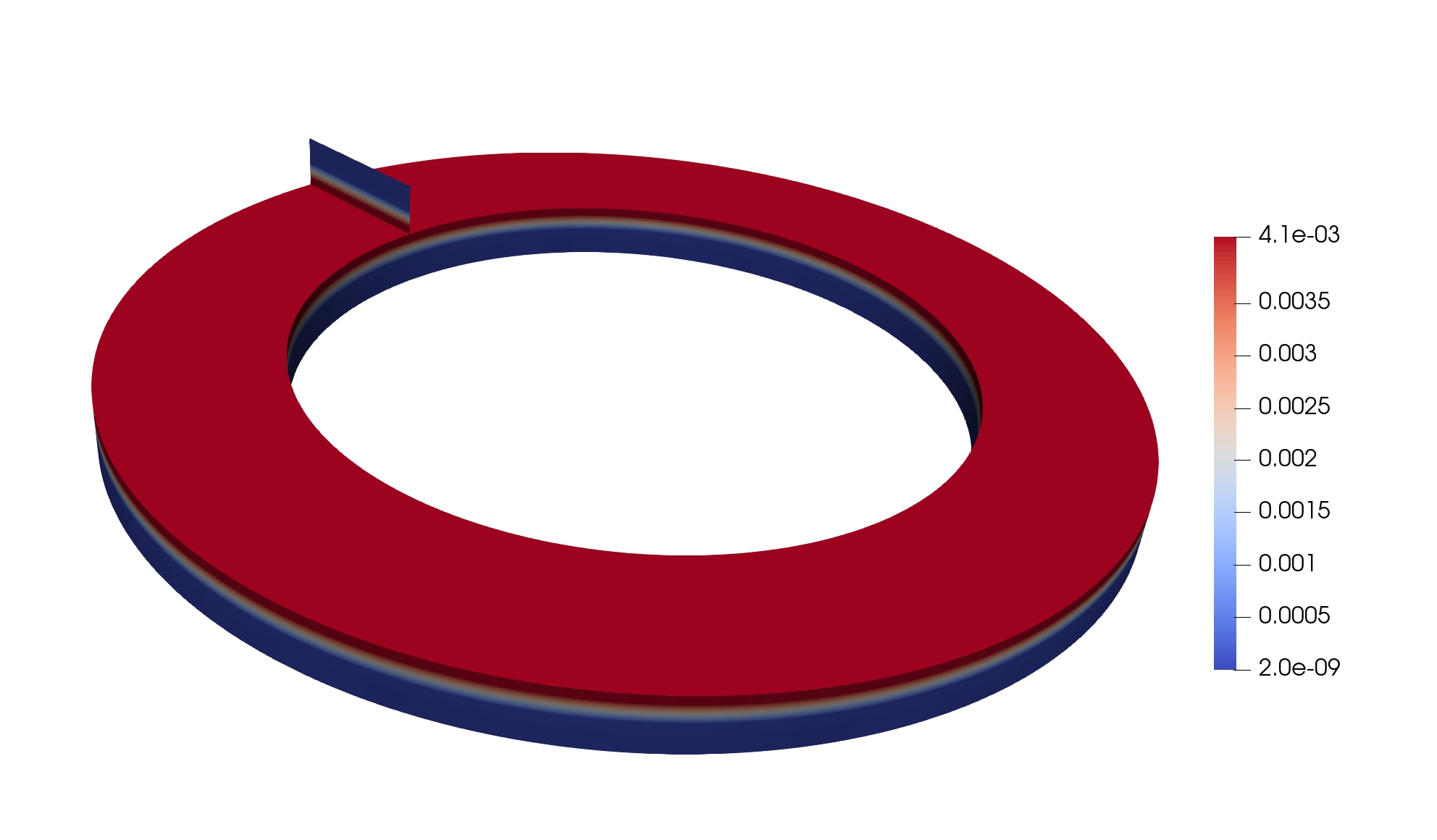}
           \includegraphics[trim = 0cm 0cm 1.7cm 1.5cm, clip, width=0.4\hsize]{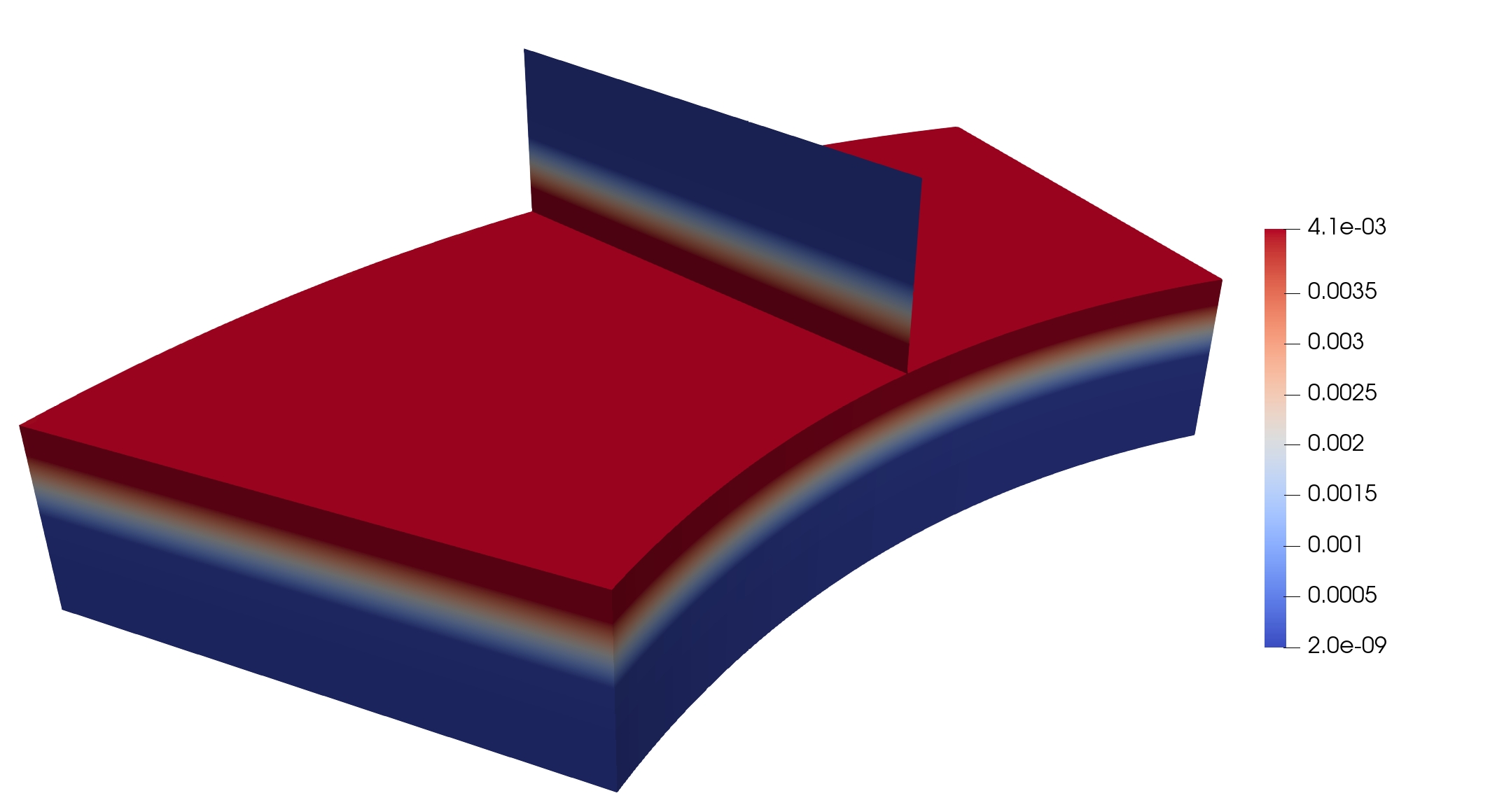}
          }
\resizebox{\hsize}{!}
          {\includegraphics[trim = 0.8cm 0cm 12cm 1.5cm, clip, width=0.35\hsize]{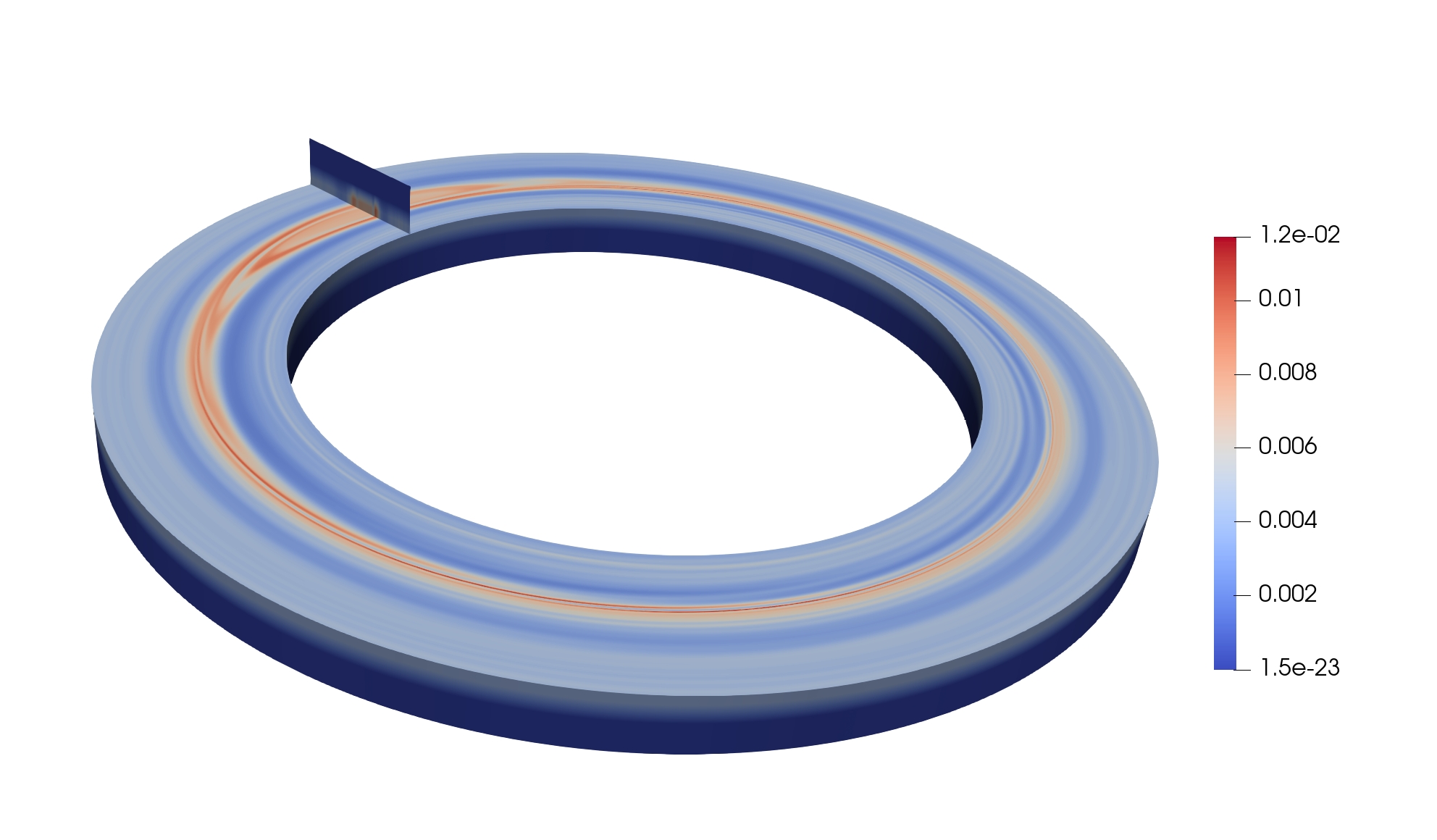}
           \includegraphics[trim = 0cm 0cm 1.7cm 1.5cm, clip, width=0.4\hsize]{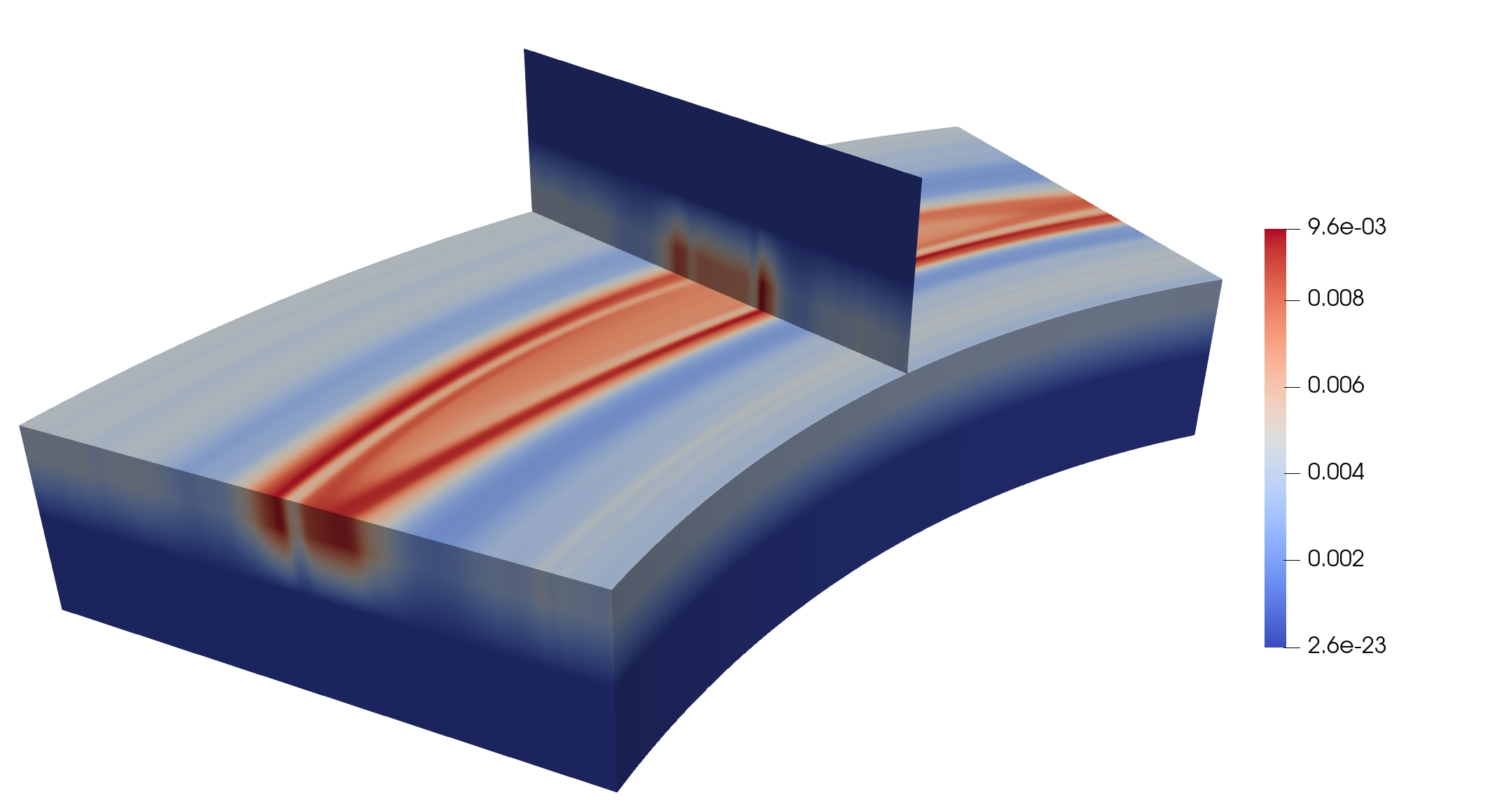}
          }
          
\resizebox{\hsize}{!}
          {\includegraphics[trim = 0.8cm 0cm 12cm 1.5cm, clip, width=0.35\hsize]{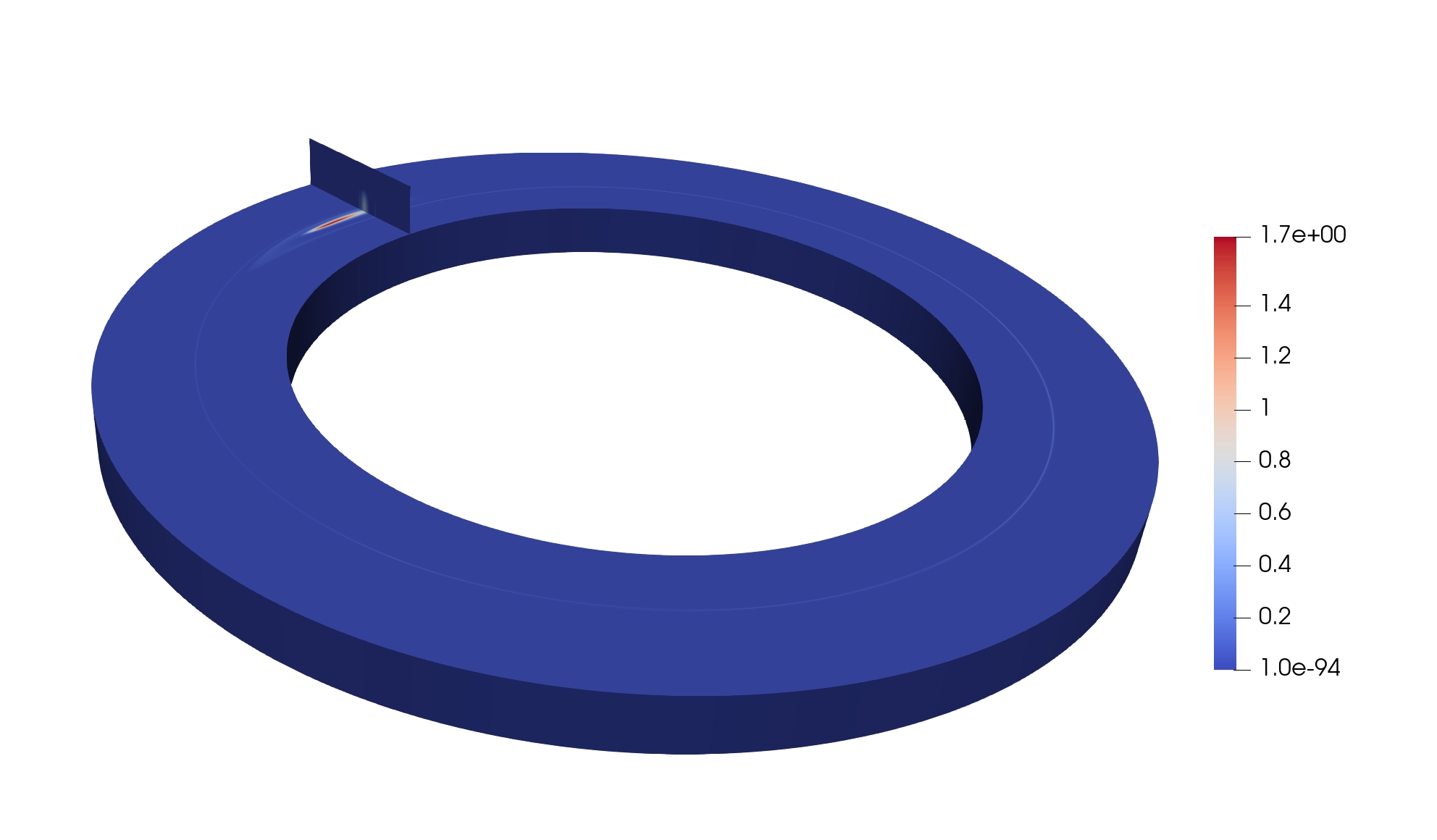}
           \includegraphics[trim = 0cm 0cm 1.7cm 1.5cm, clip, width=0.4\hsize]{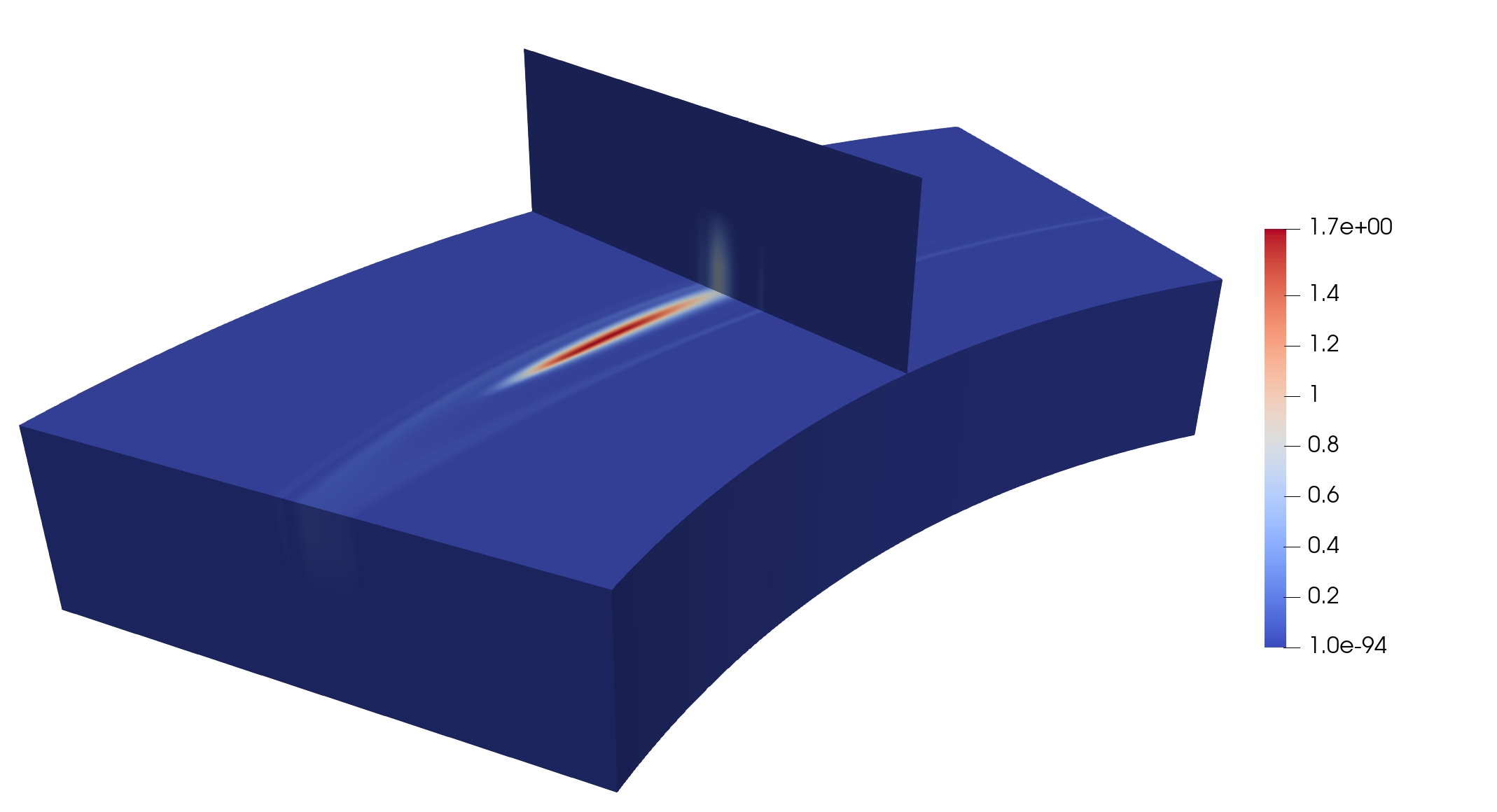}
          }
          

\caption{\textbf{Dust trapping by a 3D vortex - $\rho_{dust}(r,\theta,z)/\rho_{gas,0}(r,z=0)$} \\
\begin{tabular}{lll}
\emph{$1^{st}$ row:} $t=0\,t_0$
                     & \emph{Left:} whole disc
                     & \emph{Right:} zoom  \\
\emph{$2^{nd}$ row:} $t=2\,t_0$ 
                     & \emph{Left:} whole disc 
                     & \emph{Right:} zoom  \\
\emph{$3^{rd}$ row:} $t=30\,t_0$ 
                     & \emph{Left:} whole disc
                     & \emph{Right:} zoom  \\ 
\end{tabular}
\newline
In all these graphics we rescaled the vertical direction by a factor 20.}
\label{fig:dust trapping by 3D vortex}
\end{figure*}
%
Vortices are gaseous structures whose main interest is their strong ability to capture solid particles with nearly optimal Stokes numbers \citep{1995A&A...295L...1B,1996Icar..121..158T}.
Here we propose to find this trapping mechanism in a 3D vortex using \rossbi{}.

For this test, we use the vortex obtained after RWI saturation in the $m=4$ mode of Sect. \ref{sec: rossby wave instability}.
Fig. \ref{fig:mode 4 saturation} shows this 3D gaseous vortex used as an input for this test.
Then we introduce uniformly dust in the radial and azimutal direction but with a vertical Gaussian distribution:
\begin{equation}\label{eq:dust distribution}
\rho_{dust}(r,\theta,z)= \frac{H_g}{H_{d,0}} \, Z_{ini} \,
                         \rho_{0,gas}(r,z=0) \,
                         \exp{\left[-\left(\frac{z}{H_{d,0}}\right)^2\right]}
\end{equation}
where $H_{d,0}/H_g=0.01$ and $Z_{ini}=10^{-4}$.
We chose particles interacting with the gas only through aerodynamic forces (friction and dust back-reaction) and with a nearly optimal Stokes number ($St=0.5$).
In Fig. \ref{fig:dust trapping by 3D vortex}, the dust density is exhibited at different times.
We do not show the gaseous vortex evolution since it is nearly unaffected by the dust back-reaction.

In the early stages, we observe that dust settles in the midplane and starts to be captured by the vortex.
After $\sim 10$ orbits almost all dust lies exactly in the midplane: the settling time is consistent with the theoretical prediction $\tau_{sed}/t_0\sim\left(St+\frac{1}{St} \right)=2.5$.
Because of the absence of viscosity the dust scale vanishes (at the limit of the vertical resolution).
The gaseous vortex collects dust during the whole simulation and after 110 orbits the dust density is increased by a factor 2000 in the vortex core which is in agreement with the theoretical expectations.

\section{Next steps and possible development}\label{sec:next steps and possible improvements}

Despite a good scalability, \rossbi{} may need improvements to tackle challenging simulations, such as 2D and 3D SG at high resolutions, and at the same time, more physics options are needed to describe PPDs in a more realistic way.
Therefore, in the following sections, we discuss possible improvements to our code.

\subsection{Can GPU improve the performance of \rossbi{} ?}

%
\begin{figure}
\centering

\resizebox{\hsize}{!}
          {\includegraphics[trim = 0cm 0cm 0cm 0cm, clip, width=0.4\hsize]{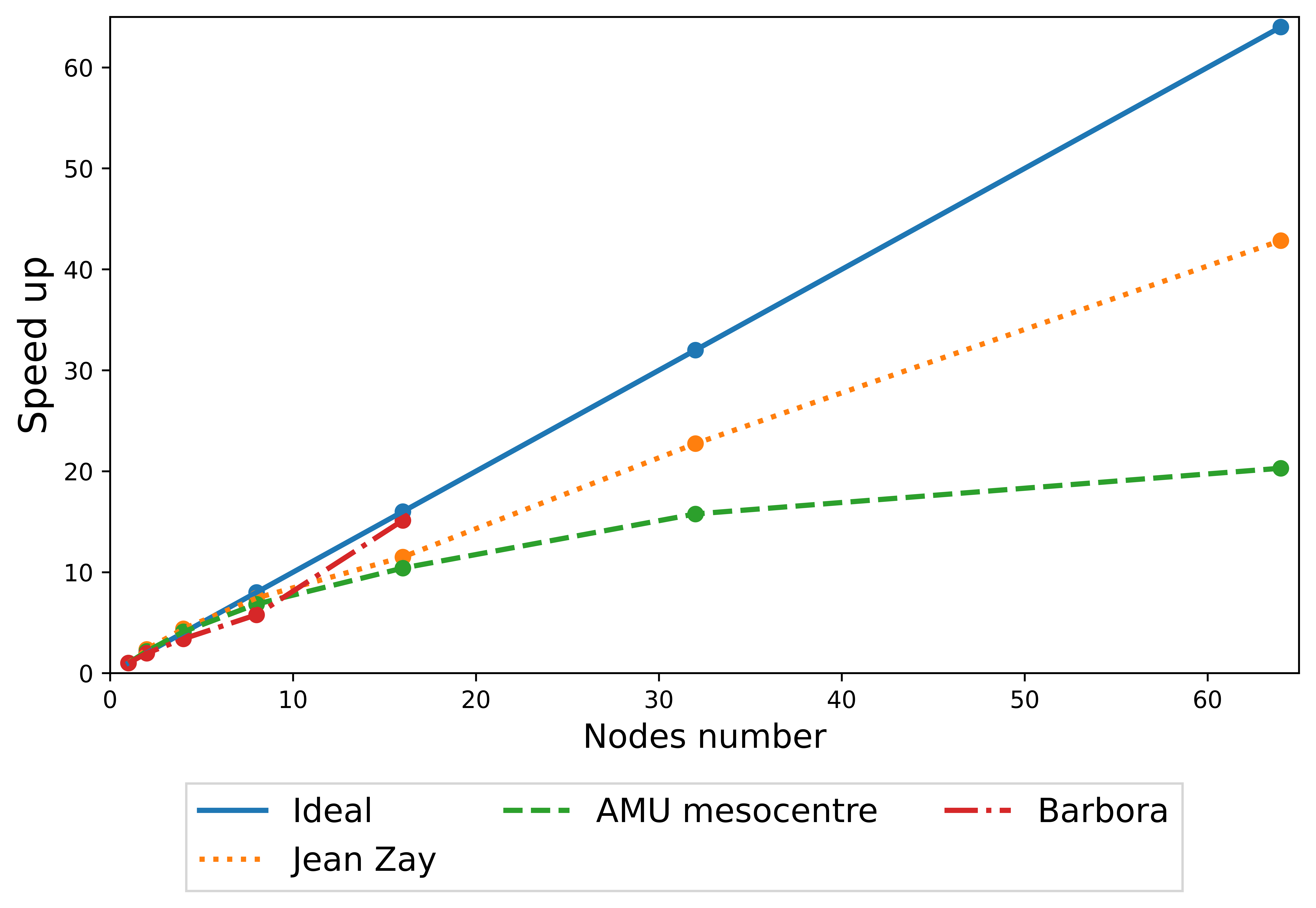}
          }
\caption{\textbf{Scalability of \rossbi{} on different supercomputers} \\
\textbf{Outline: Column Gaussian vortex evolution.} \\
\textbf{Jean Zay:} 2 Intel Cascade Lake 6248, 20 cores, 2.5 GHz (40 cores/node) \\
\textbf{Barbora:} 2 Intel Cascade Lake, 18 cores, 2.6 GHz (36 cores/node). Infiniband HDR 200 Gb/s \\
\textbf{AMU Mesocentre:} Intel\textsuperscript{\tiny\textregistered}Xeon\textsuperscript{\tiny\textregistered} Gold 6142 (SkyLake), 32 cores, 2.6 GHz. Nodes Dell PowerEdge C6420\\
}
\label{fig:scaling different machines}
\end{figure}
%

According to \citet{Gheller2015} and \citet{CUDA2020}, GPU technology is promising for solving Euler's equations.
However, the current code architecture and the observed workload on processors in tests conducted do not permit confirming yet if \rossbi{} performance would benefit notably from GPU acceleration. 
Another audit or proof of concept analysis would be necessary to elucidate this question, but this is not a current priority.
Furthermore, the parallelisation philosophy in the CUDA language is different from MPI and this would imply a significant rewriting of the code.
Alternatively, the OpenACC language could be employed for prototyping GPU acceleration since it is way less intrusive than CUDA and can potentially reach similar performances.
Finally, the metrics and scalability of the code illustrated in Table \ref{tab:my-table} and Fig. \ref{fig:scaling different machines} are satisfactory, and for the current use there is no real need to migrate to GPU.

\subsection{Optimisation of 2D self-gravity}

%
\begin{figure}
\centering

\resizebox{\hsize}{!}
          {\includegraphics[trim = 0cm 0cm 0cm 0cm, clip, width=0.4\hsize]{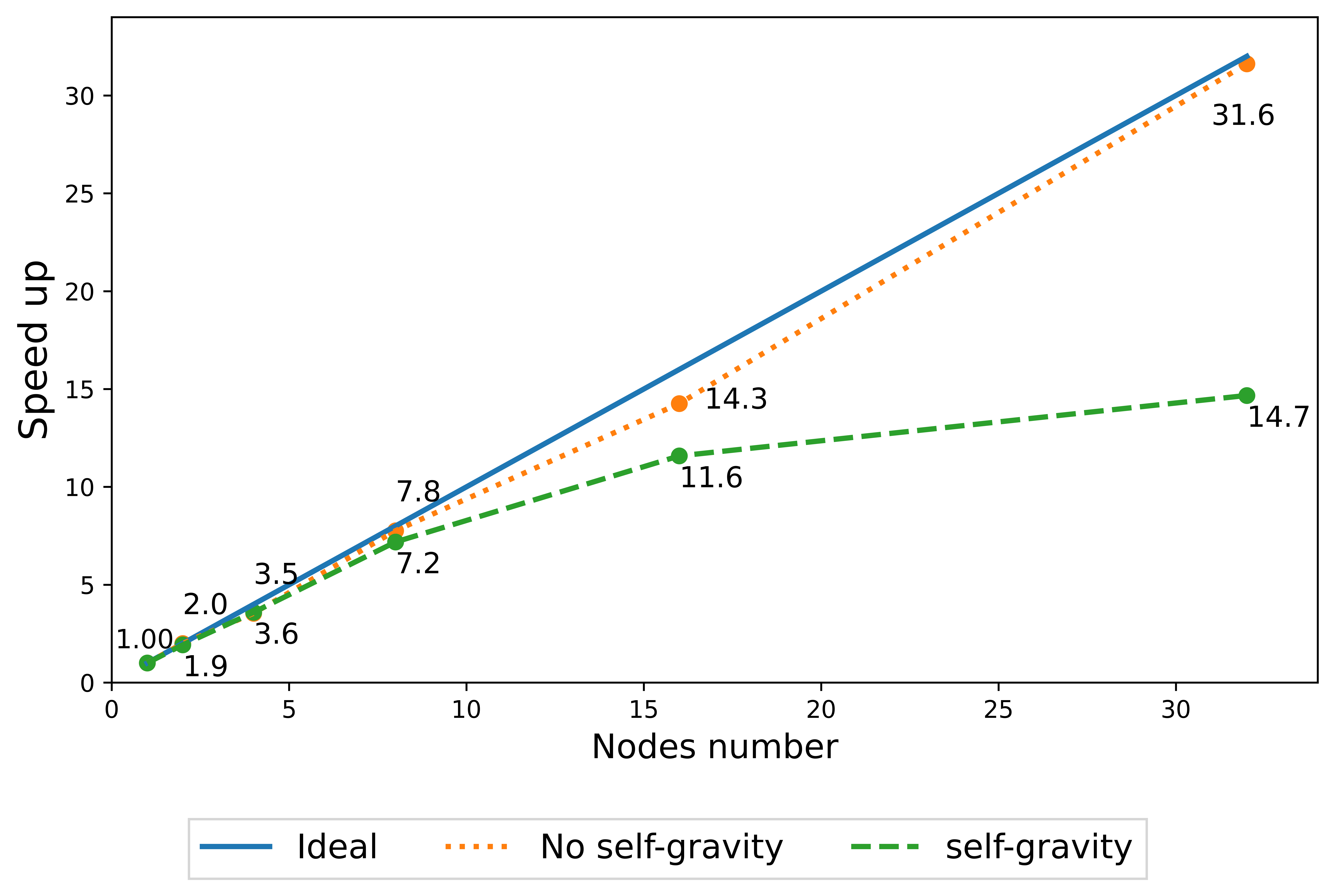}
          }
\caption{\textbf{Scalability of \rossbi{} for a 2D simulation with and without SG} \\
\textbf{Outline: Self-gravitating Gaussian vortex evolution.}                      \\
Parameters: $\chi=7$, $\Delta r/H=1.5$, $Ro=-0.14$, $h=0.05$. Resolution: $(N_r, N_\theta,N_z)=(2880,6400)$. Simulation box: $(r_{in}, r_{out})=(40,60)$, $r_0=50$. \\
\textbf{Jean Zay cluster:} 2 Intel Cascade Lake 6248, 20 cores, 2.5 GHz (namely 40 cores/node)
}
\label{fig:scaling SG vs NO_SG}
\end{figure}
%

Currently the calculation of 2D SG is based on convolution products computed thanks to fast Fourier transforms.
In Fig. \ref{fig:scaling SG vs NO_SG}, we compare the scalability of the 2D version of \rossbi{} with and without SG with respect to the number of processors.
We found that with SG, the scalability starts to show limitations from 16 nodes (576 processors), whereas such behaviour is not observed without SG.
This suggests that SG calculation is responsible for the loss of performances for a large number of nodes.
Possible improvements that permit alleviating this issue consists in performing in-place transforms, which is particularly interesting if each processor has to treat large amounts of data.
Another idea is to carrying out a single transform on a unique multidimensional array instead of doing successive transforms on multiple arrays.
This could be beneficial for bi-fluid simulations where 6 Fourier transforms need to be performed at each time step, compared to 2 for a single fluid simulation \citep{rendonbarge_2023}.

\subsection{Implementation of 3D self-gravity}\label{subsec:3D selfgravity}


SG is a rather scarce option in 3D hydrodynamic codes. 
There are only a few codes with this option, e.g.:
\href{http://pencil-code.nordita.org/}{the PENCIL Code} \citep{pencil_2021}, 
\href{https://bitbucket.org/rteyssie/ramses/src/master/}{RAMSES} \citep{ramses_2002},
\href{https://gitlab.aip.de/ziegler/NIRVANA}{NIRVANA} \citep{Ziegler2004} or 
\href{https://flash.rochester.edu/site/flashcode/user_support/}{FLASH} \citep{flash_code_2000}. 
The scarcity mainly comes from computational costs and trade offs between accuracy and computational efficiency.
In the next paragraph, we summarise few of the main methods for computing SG and identify those that could help us.

The numerical computation of the SG potential and the associated forces are handled by two main approaches.
The first one is called the difference method and is based on the resolution of the Poisson's equation:
\begin{equation}\label{eq:poisson equation}
\Delta \Phi_{sg} = 4 \pi G \rho 
\end{equation}
where $\Phi_{sg}$ is the gravitational potential.
This equation discretization results in a linear system which is usually solved using iterative multigrid methods as it was done by \citet{nirvana_2005} in NIRVANA.
Nonetheless, this requires evaluating the SG potential at the numerical domain boundaries which could be done by multipole expansion.
The second approach called integral method relies on the equivalent integral form:
\begin{equation}
\Phi_{sg}(\vr)=-G \iiint\limits_{disc} \frac{\rho(\vr')}{||\vr-\vr'||} d^3\vr' 
\end{equation}
This summation could be performed using a multipole expansion of $||\vr-\vr'||^{-1}$, which is explained in detail by \citet{SG_integral_2019}.
This second formulation is also compatible and appealing for some Adaptive Mesh Refinement (AMR) codes where one can take advantage of the tree structure:
in a crude way, the self-gravity calculation is refined in the neighbourhood of a cell or regions of strong density gradients while long distance and slow varying density regions are averaged.
This method is very efficient numerically and is used by \citet{sg_flash}.
Finally, fast Fourier methods could also be used for both formulations and different types of geometries \citep{2019ApJ...870...43M,2019ApJS..241...24M,2021ApJS..252...14K}.

\subsection{A possible multi-fluid approach}

This is a development often considered to address the complex evolution of PPDs. 
Numerical simulations with two phase codes generally describe the solid material as a single population of spherical and solid particles. 
This is of course a major draw-back which biases the estimate of dust-gas friction and, consequently, the description of the disc evolution. 
A natural possibility to overcome this problem is to use a multi-fluid approach as, for example, the ones implemented by \citet{Hanasz2010}, \citet{Benitez2019} and \citet{Huang2022}.

However, in a view of a more realistic evolution, it is also important to describe the exchanges of mass and energy between the various fluids of the dispersed phase and between the two phases, gas and solids. 
This refers to the formation of planetesimals, a problem in which binary collisions and gas drag must be taken into account and result in coagulation/fragmentation or condensation/evaporation processes \citep{Weiden1980,  Voelk1980, Weiden1993, Brauer2008, Birnstiel2011}. 
An appropriate way to implement a multi-fluid approach could be to use a sectional method \citep{Laurent2001,Laurent2002}, which is well known in the context of the dispersed droplet evolution \citep{Dufour2005} and can be interpreted in terms of the kinetic theory of gases \citep{Tambour1985, Laurent2001}. 
The advantages of this method are to ensure conservation of moment transfer between the various fluids and, possibly, to rely on the large developments of the numerical simulation of reactive flows \citep{Drui2017}.

\subsection{Other possible improvements}\label{subsec:other improvements}

In order to explain the angular momentum transport in PPDs, it is common to associate local turbulence, generated in various regions of the disc, with a pseudo-viscosity \citep{1973A&A....24..337S}.
Promising mechanisms generating turbulence include the magneto rotational instability (MRI) \citep{1991_balbus} and the VSI \citep{nelson_gressel_2013, 2014A&A...572A..77S}, among others.
Such mechanisms encourage to include viscosity in the \rossbi{} code for a realistic description of PPD.
However, accounting viscous effects modifies the nature of the equations to be solved: Euler's equations are hyperbolic whereas Navier-Stokes equations are parabolic.
The presence of diffusion causes the Riemann solver to be inadequate for solving the Navier-Stokes equations, forcing an alternative numerical strategy.
A possible solution could be to use an explicit Runge-Kutta-Legendre method \citep{2014_meyer}.

The confrontation of numerical models with dust thermal emission \citep{2015PASP..127..961A} is key since it allows discriminating between physical processes operating on PPDs. 
Such comparison is made possible thanks to dust radiative transfer codes, see \citep[and references therein]{Steinacker}.
Therefore, an important step is to include radiative transfer in the post-processing pipeline from another existing code.
We think that the best option is to use a free, robust, documented, and up to date existing code: the most attractive solutions seem to be \href{https://github.com/dullemond/radmc3d-2.0}{RADMC-3D 2.0} or \href{https://ipag.osug.fr/~pintec/mcfost/docs/html/overview.html}{MCFOST}.


\section{Conclusions}\label{sec:conclusion}

In this paper, we presented the new \rossbi{} code specifically designed for PPDs studies at 2D and 3D.
This FORTRAN 90 code solves the fully compressible inviscid Euler, continuity and energy conservation equations in polar coordinates for an ideal gas orbiting a central object.
Solid particles are treated as a pressureless fluid and interact with the gas through aerodynamic forces in the Epstein or Stokes regimes.
In 2D, particles can also interact with gas via SG.
The main features implemented in this code are:
\renewcommand\labelitemi{{\boldmath$\cdot$}}
\newline
\newline
\centerline{\textbf{Numerical methods}}
\begin{itemize}
\item Finite volume scheme adapted to non-linear conservation laws and naturally to computational fluid dynamics problems
\item Second order Runge Kutta method for the time integration
\item An exact Riemann solver 
\end{itemize}

\centerline{\textbf{Scheme improvement and stability}}
\begin{itemize}
\item Sheared BC at radial boundaries for avoiding waves reflection and instabilities. Outflow BC in the vertical extents
\item[\boldmath$\cdot$] A CFL-like condition respected in the whole numerical domain which ensures time integration stability
\item A balanced scheme adapted to PPDs
\item A parabolic interpolation based on deviations to the equilibrium solution
\item A MinMod flux limiter allowing to handle drastic density gradients
\end{itemize}

\centerline{\textbf{An optimised parallelism}}
\begin{itemize}
\item A 3D domain decomposition for 3D calculations
\item An optimised code thanks to POP audit and good MPI practices
\item A satisfactory scalability up to 1280 processors (2D with SG) and 2560 processors (2D and 3D without SG)
\end{itemize}
We performed test cases and we retrieved expected results, which validated \rossbi{} reliability.
As future improvements, we propose the following implementations:
\begin{itemize}
\item Viscosity at 2D and 3D
\item A 3D SG option based on the integral method and Fast Fourier Transforms
\item Output files compatible with RADMC-3D 2.0 or MCFOST to include radiative transfer
\end{itemize}
For 3D calculations, the code is accompanied by a visualisation tool based on Paraview.
Finally, \rossbi{} is open source, released under the terms of the CeCILL2 Licence, and accessible from its public repository: \href{git@gitlab.lam.fr:srendon/rossbi3d.git}{git@gitlab.lam.fr:srendon/rossbi3d.git}.

\begin{acknowledgements}
S.R.R acknowledges Clément Baruteau for discussions and suggestions for a future coding of SG at 3D, Jean M. Favre for writing the Python programmable source used for 3D data visualisation in Paraview, and Jean-Charles Lambert for his technical support during the whole coding phase of the 3D version and optimisation of the 2D version.
We would like to warmly acknowledge Stéphane Le Dizès for fruitfull discussions during the preparation of the paper and also for his help to find funding for part of the project. 
P.B thanks Serguei Rodionov at LAM for his help in initiating the first systematic treatments of the data using Python subroutines.
This work was granted access to the HPC resources of IDRIS under the allocation A0090411982 made by GENCI. 
Centre de Calcul Intensif d’Aix-Marseille is acknowledged for granting access to its high performance computing resources.
This research has made use of computing facilities operated by CeSAM data center at LAM, Marseille, France.
This work was supported by the POP2 project - the European Union’s Horizon 2020 research and innovation programme under grant agreement No. 824080 and by the Ministry of Education, Youth and Sports of the Czech Republic through the e-INFRA CZ (ID:90140).
Co-funded by the European Union (ERC, Epoch-of-Taurus, 101043302). 
Views and opinions expressed are however those of the author(s) only and do not necessarily reflect those of the European Union or the European Research Council. 
Neither the European Union nor the granting authority can be held responsible for them.
\end{acknowledgements}

%
%

\bibliographystyle{aa}
\bibliography{bibliography}

\begin{appendix}

\section{Stationary gas flow at equilibrium}

Assuming the flow of gas is axisymmetric, in Keplerian equilibrium in the radial direction, in hydrostatic equilibrium in the vertical direction and inviscid, we get the equilibrium solution:
\begin{equation}\label{Eq: equilibrium solution}
\left\{\begin{array}{ccl}
T_0(r)       & = & T_0 r^{\beta_T}                                    \\ [8pt]
\rho_0(r,z)  & = & \displaystyle \frac{\sigma_0(r)}{\sqrt{\pi} H(r)} \ 
                  \displaystyle\exp{\left( \frac{G M_{\odot}}{c_s^2(r)}
                  \left(\frac{1}{\sqrt{r^2+z^2}} 
                  - \frac{1}{r}\right) \right)}                       \\ [10pt]
P_0(r,z)     & = & \displaystyle \frac{k_B}{\mu m_u} \rho(r,z) \, T(r)\\ [8pt]
v_{r,0}(r,z) & = & 0                                                  \\ [6pt]
v_{\theta,0}(r,z) 
             & = & \displaystyle\sqrt{- r f_{star,r} +
                   \frac{r}{\rho} \frac{\partial P}{\partial r}}      \\ [10pt]
v_{z,0}(r,z) & = & \displaystyle 0                               
\end{array}\right.
\end{equation}
where : 
\begin{equation}
\left\{\begin{array}{ccl}
\sigma_0(r) & = & \sigma_0 r^{\beta_\sigma}                        \\ [8pt] 
H(r)        & = & \displaystyle \sqrt{\frac{2}{\gamma}} 
                  \displaystyle \frac{c_s(r)}{\Omega_K(r)} \\ [4pt]
f_{star,r}(r,z) 
            & = & \displaystyle -\frac{G M_{\odot} r}{(r^2+z^2)^{\frac{3}{2}}} 
                                                                   \\ [4pt]
\end{array}\right.
\end{equation}

%

\section{Conservative form variables}\label{app:conservative form variables}

Below we present the Euler variables, flux terms and source terms. 
\newline\newline
\centerline{\textbf{Euler Variables}}
\begin{equation}
E=\left(
\begin{array}{ccc}
\rho u    \\
\rho v    \\
\rho w    \\
\mathcal{E} \\
\rho
\end{array}\right)
\end{equation}

\centerline{\textbf{Flux terms}}
\begin{equation}
\vec{F}=\left(
\begin{array}{lclcl}
\left(\rho u^2 + P \right)\Vec{e}_r  
     & + & \rho u v \vec{e}_\theta 
     & + &\rho u w \vec{e}_z                              \\[0.5ex]
\rho u v\Vec{e}_r 
     & + &\left(\rho v^2 + P \right) \vec{e}_\theta 
     & + &\rho v w \vec{e}_z                              \\[0.5ex]
\rho u w\Vec{e}_r 
     & + &\rho v w \vec{e}_\theta 
     & + &\left(\rho w^2 + P \right) \vec{e}_z            \\[0.5ex]
\rho \left( \mathcal{E} + P \right)  u\vec{e}_r 
     & + &\rho \left( \mathcal{E} + P \right) v\vec{e}_\theta 
     & + &\rho \left( \mathcal{E} + P \right) w\vec{e}_z            \\[0.5ex]
\rho  u\vec{e}_r 
     & + &\rho v\vec{e}_\theta 
     & + &\rho w\vec{e}_z 
\end{array}\right)
\end{equation}

\centerline{\textbf{Source terms}} 
\begin{equation}
\vec{S}=\left(
\begin{array}{lcc}
\displaystyle S_1^1 + S_1^2 + S_1^3                    \\[2ex]
\displaystyle -\frac{\rho uv}{r} +  \rho f_{V,\theta}  \\[2ex]
\displaystyle \rho f_{V,z}                             \\[2ex]
\displaystyle S_4^1 + S_4^2 + S_4^3 + S_4^4            \\[2ex]
\mathcal{S}_\rho  
\end{array}\right)
\end{equation}
with:
\begin{equation}
\left\{
\begin{array}{lccc}
\displaystyle S_1^1 = \frac{\rho v^2}{r}   & S_1^2 = \rho f_{V,r}     & S_1^3 = \displaystyle\frac{P}{r} & \\[2ex]
\displaystyle S_4^1 = \rho u f_{V,r}       & S_4^2 = \rho v f_{V,\theta}   & S_4^3 = \rho w f_{V,z}  & S_4^4 = \mathcal{S}_\mathcal{E}
\end{array}\right.
\end{equation}
where $f_V$ is a volume force.

\section{Data treatment and visualisation tools}

\rossbi{} produces data for meshes containing up to few millions of points which require efficient visualisation tools. 
At the early stages of development of this code, the interactive visualisation program \href{https://projets.lam.fr/projects/glnemo2/wiki}{GLnemo2}\footnote{https://projets.lam.fr/projects/glnemo2/wiki} was used. 
Despite many advantages it was later decided to resort to another visualisation software.
The decision was based mostly on the need to make visualisation runs remotely and in parallel but also with the need to display up to 15 different physical variables, which are not native in GLnemo2.

In order to satisfy above requirements, it was decided to use \href{https://www.paraview.org/}{Paraview} \citep{Paraview}.
This visualisation application allows data to be visualised remotely using the GPU/CPU resources of computing clusters, which improves rendering since the graphic card of the host station is not used.
Currently, the visualisation works thanks to a \textit{programmable source} based on a \href{https://www.python.org/}{Python} script.
In order to fully benefit from Paraview capabilities, it would be necessary in the future to write the output files of the \rossbi{} code as \href{https://vtk.org/}{VTK} files. \\

\end{appendix}

\end{document}